\documentclass[%
 reprint,
 amsmath,amssymb,
 aps,
]{revtex4-2}
\usepackage[utf8]{inputenc} 
\DeclareUnicodeCharacter{0229}{\c{e}}
\DeclareUnicodeCharacter{2013}{--}
\usepackage[T1]{fontenc}
\usepackage{graphicx}
\usepackage{dcolumn}
\usepackage{bm}
\usepackage{here}
\usepackage{physics}
\usepackage{amsmath}
\usepackage{amssymb}
\usepackage{amsfonts}
\usepackage{ascmac}
\usepackage{xr}

\begin{document}
\preprint{APS/123-QED}
\title{Decomposing Thermodynamic Dissipation of Linear Langevin Systems via Oscillatory Modes and Its Application to Neural Dynamics}
\author{Daiki Sekizawa}
\email{sekizawa-daiki963@g.ecc.u-tokyo.ac.jp}
\affiliation{
Department of General Systems Studies, The University of Tokyo, 3-8-1 Komaba, Meguro-ku, Tokyo 153-8902, Japan}
\author{Sosuke Ito}
\email{s-sosuke.ito@g.ecc.u-tokyo.ac.jp}
\affiliation{Department of Physics, The University of Tokyo, 7-3-1 Hongo, Bunkyo-ku, Tokyo 113-0033, Japan}
\affiliation{Universal Biology Institute, The University of Tokyo, 7-3-1 Hongo, Bunkyo-ku, Tokyo 113-0033, Japan}
\author{Masafumi Oizumi}
\email{c-oizumi@g.ecc.u-tokyo.ac.jp}
\affiliation{Department of General Systems Studies, The University of Tokyo, 3-8-1 Komaba, Meguro-ku, Tokyo 153-8902, Japan}
\date{\today}

\begin{abstract}
Recent developments in stochastic thermodynamics have elucidated various relations between the entropy production rate (thermodynamic dissipation) and the physical limits of information processing in nonequilibrium dynamical systems. These findings have been actively utilized and have opened new perspectives in the analysis of real biological systems. In neuroscience also, the importance of quantifying entropy production has attracted increasing attention as a means to understand the properties of information processing in the brain. However, the relationship between entropy production rate and oscillations, which is prevalent in many biological systems, is unclear. For example, neural oscillations, such as delta, theta, and alpha waves play crucial roles in brain information processing. Here, we derive a novel decomposition of the entropy production rate of linear Langevin systems. We show that one of the components of the entropy production rate, called the housekeeping entropy production rate, can be decomposed into independent positive contributions from oscillatory modes. Our decomposition enables us to calculate the contribution of oscillatory modes to the housekeeping entropy production rate. In addition, when the noise matrix of the Langevin equation is diagonal, the contribution of each oscillatory mode is further decomposed into the contribution of each element of the system. To demonstrate the utility of our decomposition, we applied our decomposition to an electrocorticography (ECoG) dataset recorded during awake and anesthetized conditions in monkeys, wherein the properties of oscillations change drastically. We showed the consistent trends across different monkeys, i.e. the contribution of oscillatory modes from the delta band were larger in the anesthetized condition than in the awake condition, while those from the higher frequency bands, such as the theta band, were smaller. These results allow us to interpret the change in neural oscillation in terms of stochastic thermodynamics and the physical limit of information processing.

\end{abstract}

\maketitle

\section{Introduction}

Oscillatory phenomena play crucial roles in various biological systems, such as cellular oscillations~\cite{lodish2008molecular} and the rhythmic contractions of the heart~\cite{qu2014nonlinear}, and oscillations of neural activity in the brain. Neural oscillations, which we focus on this study, are integral to brain functions such as information processing, memory, and consciousness. Brain waves like alpha, beta, and gamma waves exhibit distinct oscillatory patterns corresponding to different cognitive states and activities~\cite{Buzsaki2006-se}. Additionally, abnormal synchronous oscillations, such as those occurring during epileptic seizures, can disrupt normal brain functions~\cite{engel2008epilepsy}. These examples underscore the universal presence and significance of oscillatory dynamics across these fields.

In understanding these non-equilibrium biological systems, the entropy production rate in stochastic thermodynamics~\cite{sekimoto2010stochastic, van2010three, seifert2012stochastic} has opened new research directions~\cite{gnesotto2018broken}. The entropy production rate is a thermodynamic dissipation which quantifies the thermodynamic temporal irreversibility or degree of non-equilibrium in a system's dynamics. Moreover, stochastic thermodynamics has elucidated relationships between information processing and entropy production rate~\cite{Parrondo2015-wo, seifert2019stochastic, ito2023geometric}. So far, several relationships between the entropy production rate and the physical limits of information processing have been discovered. For example, the entropy production rate can determine various theoretical limits of information processing, such as the speed limits of information processing~\cite{aurell2012refined,chen2019stochastic,van2021geometrical, Nakazato2021-do,yoshimura2023housekeeping}, the limit of accuracy of information processing~\cite{barato2015thermodynamic, dechant2018current, seifert2019stochastic, horowitz2020thermodynamic}, and the limit of the performance of biochemical adaptation~\cite{lan2012energy, sartori2014thermodynamic,barato2014efficiency, ito2015maxwell}. Methods for calculating the entropy production rate in large-scale complex networks have been developed \cite{aguilera2021unifying}, and relationships between the entropy production rate and criticality \cite{munoz2018colloquium, o2022critical}, which is suggested to play an important role in information processing, are also being explored \cite{aguilera2023nonequilibrium}. 

However, despite the utility of the entropy production rate to the understanding of biological systems, its relationship to oscillations, which are ubiquitous phenomena in these fields, remains elusive. In this paper, we first derive a relation between the entropy production rate and oscillation. Then, by applying the novel relation to neural activity data, we investigate how differently oscillatory phenomena contribute to the entropy production rate in different brain conditions, such as awake and anesthetized conditions.

In particular, we show theoretically that the housekeeping entropy production rate can be decomposed into independent positive contributions from oscillatory modes, inspired by dynamic mode decomposition~\cite{schmid2010dynamic,kutz2016dynamic}. Our decomposition is widely applicable to linear Langevin systems. The housekeeping entropy production rate is one component from the geometric decomposition of the entropy production rate~\cite{Maes2014-pk, Dechant2022-gt}, which is explained in Section \ref{section_ex_hk}. Dynamic mode decomposition is a method used to decompose multi-dimensional time series data into several oscillatory modes, and is widely used in data analysis. Dynamic mode decomposition has advantages over the classical Fourier transformation in data analysis, especially because it can capture coherent oscillatory patterns across dimensions. Our decomposition of the housekeeping entropy production rate into oscillatory modes is based on dynamic mode decomposition. In our results, we newly identified the contributions to the housekeeping entropy production rate of each dynamic mode of the housekeeping part of the dynamics. The contributions of the mode are written as the product of the square of the frequency and the oscillatory power of the multi-dimensional modes.

Additionally, to gain insight we illustrate our decomposition with two toy examples. In example 1, we explain our decomposition using a simple analytically tractable system which has only one oscillatory mode. In example 2, we explain our decomposition using a system which generates multiple oscillatory modes.

Further, to demonstrate the utility of our framework for quantifying the relation between the entropy production rate and oscillation in the brain, we applied our mode decomposition to real neural data. In the brain, oscillatory phenomena such as delta, theta, alpha, beta, and gamma waves are universally observed. We sought to compare the contributions of each oscillatory mode to the housekeeping entropy production rate across multiple conditions with different oscillatory properties. For this purpose, we chose to analyze a monkey ECoG dataset during different awake conditions (eyes-closed and eyes-opened conditions) and anesthetized conditions~\cite{Nagasaka2011-gk} as a typical example, because it is well known that the amplitude of alpha waves increases during the awake eyes-closed condition while the amplitude of delta waves increases during the anesthetized condition~\cite{Miyasaka1968-zz}. We robustly observed that the contribution of oscillatory modes from the delta band (0.5-4Hz) was larger in the anesthetized condition, whereas the contribution of oscillatory modes from the theta band (4-7Hz) were smaller in the awake eyes-closed condition.

These results may allow thermodynamic interpretation of oscillatory phenomena, which are ubiquitous in biological and chemical systems. This may in future reveal oscillation-dependent thermodynamic limits of information processing in biological systems, including the brain.

\section{Background}
\subsection{Stochastic thermodynamics for Langevin equation and Fokker--Planck equation}
Here we explain the setup of our study. We consider the multidimensional Langevin equation for the dynamics of the state in $d$-dimensional space $\bm{x}_t \in \mathbb{R}^{d}$ at time $t$:
\begin{align}
 d\bm{x}_t=\mathit{D}_t\bm{f}_t(\bm{x}_t)dt+\sqrt{2}\mathit{G}_td\bm{B}_t \label{eq:Langevin}
\end{align}
where  $d\boldsymbol{x}_t$ is the increment of the state, $dt$ stands for the infinitesimal time interval,  $\bm{f}_t(\bm{x}_t)$ is the force at state $\bm{x}_t$ and time $t$, $\mathit{G}_t$ is the $d \times d$ matrix representing the strength of the noise at time $t$, and $D_t$ is the diffusion matrix at time $t$ defined as
$\mathit{D}_t=\mathit{G}_t\mathit{G}_t^\top$. We assume that $\mathit{G}_t$ is a regular matrix and the inverse matrix $G_t^{-1}$ can be introduced. 
The symbol $^\top$ stands for the transpose of the matrix. $d\boldsymbol{B}_t$ denotes a standard $d$-dimensional Brownian motion, which is defined as the Wiener process satisfying $\mathbb{E}[d\boldsymbol{B}_t]= \boldsymbol{0}$ and $\mathbb{E}[d \boldsymbol{B}_{t}d\boldsymbol{B}_{s}^\top]=\delta(t-s) I dt$ where $\mathbb{E}[\cdot]$ stands for the expected value and $I$ is the identity matrix.
We remark that the diffusion matrix $\mathit{D}_t$ is symmetric $\mathit{D}_t= \mathit{D}_t^{\top}$ and positive definite. The diffusion matrix $\mathit{D}_t$ determines the intensity of the noise given by the covariance of the noise $\mathbb{E}[(\sqrt{2}\mathit{G}_td\bm{B}_t) (\sqrt{2}\mathit{G}_td\bm{B}_s)^{\top}] = 2 D_t \delta(t-s)$.
The notation of placing $D_t$ in front of $\bm{f}_t(\bm{x}_t)$ in the Langevin equation is used to simplify notation of the excess and housekeeping local mean velocity, which will be introduced in Section II-B. 

The Langevin equation can be reformulated using the following Fokker--Planck equation:
\begin{align}
    \pdv{p_t(\bm{x})}{t}&=-\nabla \cdot \qty[\bm{\nu}_t(\bm{x})p_t(\bm{x})] \label{eq:Fokker--Planck}\\
    \bm{\nu}_t(\bm{x})&=\mathit{D}_t(\bm{f}_t(\bm{x})-\nabla\ln p_t(\bm{x}))\label{eq:local_mean_velocity}
\end{align}
The Fokker--Planck equation is a deterministic equation for the probability distribution that describes the temporal evolution of the probability distribution $p_t(\bm{x})$. The velocity field $\bm{\nu}_t(\bm{x})$ is called the local mean velocity. When $p_t$ does not change with time, the system is said to be in the steady state.

In stochastic thermodynamics, the entropy production rate $\sigma_t$ is the thermodynamic cost representing the irreversibility of the system \cite{seifert2012stochastic} and is known to determine various information processing limits \cite{barato2015thermodynamic, horowitz2020thermodynamic, aurell2012refined, ito2023geometric}. We now assume that the parity of the state $\bm{x}_t$ is even. This assumption means that $\bm{x}_t$ should not be odd variables such as velocity, and the sign of $\bm{x}_t$ cannot be changed under the transformation of time reversal.
The entropy production rate $\sigma_t$ is a non-negative quantity, and its non-negativity is regarded as the second law of thermodynamics \cite{seifert2012stochastic}. For the Fokker-Planck equation (\ref{eq:Fokker--Planck}), the entropy production rate is defined as 
\begin{align}
\sigma_t &=\langle (\bm{\nu}_t)^\top \mathit{D}_t^{-1}\bm{\nu}_t\rangle_t \nonumber \\
&= \int d\boldsymbol{x} \qty(\bm{\nu}_t(\bm{x}))^\top \mathit{D}_t^{-1}\bm{\nu}_t(\bm{x}) p_t(\boldsymbol{x}) \label{eq:EP_nu},
\end{align}
where $D_t^{-1} = (G^{-1}_t)^{\top} G^{-1}_t$ and $\langle \cdots \rangle_t = \int  d\boldsymbol{x} p_t(\boldsymbol{x}) \cdots$ stands for an expected value at time $t$. 

By considering the path probability distributions, this entropy production rate is generally given by an informational measure called the Kullback--Leibler (KL) divergence \cite{kawai2007dissipation} (see also Appendix~\ref{eq:appendixA} for the detailed derivation)
\begin{align}
\sigma_t dt &= \mathrm{D_{KL}}[\mathbb{P}_{\rm
F}\|\mathbb{P}_{\rm B}] \nonumber\\
&= \int d\boldsymbol{x}_t d \boldsymbol{x}_{t+dt} \mathbb{P}_{\rm F}(\boldsymbol{x}_{t+dt}, \boldsymbol{x}_t) \ln \frac{\mathbb{P}_{\rm F}(\boldsymbol{x}_{t+dt}, \boldsymbol{x}_t)}{\mathbb{P}_{\rm B}(\boldsymbol{x}_{t+dt}, \boldsymbol{x}_t)}, \label{eq:kldivergence}
\end{align}
where $\mathbb{P}_{\rm F}(\boldsymbol{x}_{t+dt}, \boldsymbol{x}_t)$ and $\mathbb{P}_{\rm B}(\boldsymbol{x}_{t+dt}, \boldsymbol{x}_t)$ are the forward and backward path probability distributions, respectively. The forward and backward path probability distributions $\mathbb{P}_{\rm F}$ and $\mathbb{P}_{\rm B}$ for the Langevin equation [Eq.~(\ref{eq:Langevin})] are defined as
\begin{align}
\mathbb{P}_{\rm F}(\boldsymbol{x}_{t+dt}, \boldsymbol{x}_t) &=\mathbb{T} (\boldsymbol{x}_{t+dt} | \boldsymbol{x}_t) p_t (\boldsymbol{x}_t), \\
\mathbb{P}_{\rm B}(\boldsymbol{x}_{t+dt}, \boldsymbol{x}_t) &=\mathbb{T} (\boldsymbol{x}_t | \boldsymbol{x}_{t+dt}) p_{t+dt} (\boldsymbol{x}_{t+dt}),
\end{align}
where the transition probability $\mathbb{T} (\boldsymbol{x}_{t+dt} | \boldsymbol{x}_t)$ is given by the expression with the Onsager--Machlup function~\cite{risken1996fokker},
\begin{align}
\mathbb{T} (\boldsymbol{x}_{t+dt} | \boldsymbol{x}_t) =\frac{\exp \left(- \frac{(\dot{\boldsymbol{x}}_t-D_t \boldsymbol{f}_t (\boldsymbol{x}_t))^{\top}D_t^{-1} (\dot{\boldsymbol{x}}_t-D_t \boldsymbol{f}_t (\boldsymbol{x}_t) ) dt} {4} \right)}{(4\pi dt)^{d/2} \sqrt{ \det D_t}}. \label{eq:condprob}
\end{align}
where $\dot{\boldsymbol{x}}_t = [\boldsymbol{x}_{t+dt} - \boldsymbol{x}_t]/dt$.
 The result [Eq.~(\ref{eq:kldivergence})] is a straightforward consequence of the fluctuation theorem~\cite{chernyak2006path, seifert2012stochastic}. Mathematically, $\mathrm{D_{KL}}[\mathbb{P}_{\rm F}\|\mathbb{P}_{\rm B}]$ quantifies the difference between the forward path probability distribution $\mathbb{P}_{\rm F}$ and the backward path probability distribution $\mathbb{P}_{\rm B}$. Because $\mathbb{P}_{\rm F} = \mathbb{P}_{\rm B}$ means reversibility of the dynamics, the entropy production rate 
$\sigma_t= \mathrm{D_{KL}}[\mathbb{P}_{\rm
F}\|\mathbb{P}_{\rm B}]/dt$
is regarded as an informational measure of irreversibility. 

\subsection{Geometric decomposition of entropy production rate into excess and housekeeping parts}\label{section_ex_hk}

\begin{figure}[t]
    \centering
    \includegraphics[width=\linewidth]{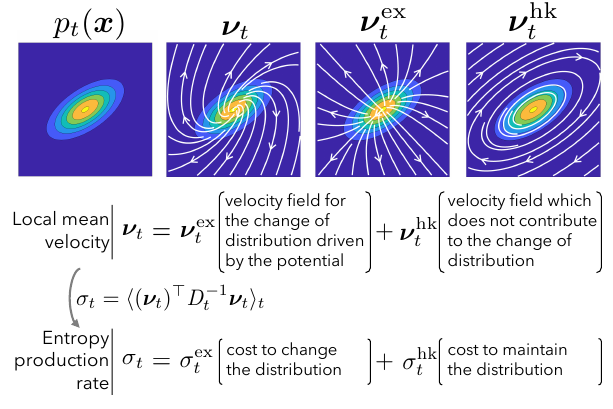}
    \caption{  
    Schematic illustration of the geometric decomposition of the entropy production rate~\cite{Dechant2022-gt, yoshimura2023housekeeping}. 
    Top: Illustration of the probability distribution $p_t(\bm{x})$ and the decomposition of the local mean velocity $\bm{\nu}_t$ in two-dimensional system. We can observe that the excess part of the local mean velocity $\bm{\nu}_t^\mathrm{ex}$ changes the probability distribution $p_t(\bm{x})$, while the housekeeping part of the local mean velocity $\bm{\nu}_t^\mathrm{hk}$ does not change the distribution.
    Bottom: The local mean velocity $\bm{\nu}_t$ can be decomposed into the excess part $\bm{\nu}_t^\mathrm{ex}$ and the housekeeping part $\bm{\nu}_t^\mathrm{hk}$ [Eq.~(\ref{eq:MN_decomposition_nu})]. The entropy production rates corresponding to each local mean velocity are the expected values of the $L^2$-norm of them [Eqs.~(\ref{eq:EP_nu}), (\ref{eq:EP_ex}) and (\ref{eq:EP_hk})]. The orthogonality of the excess and housekeeping parts of the local mean velocity with respect to the expected value [Eq.~(\ref{eq:orthogonal})] gives a geometric decomposition of the entropy production rate into the excess and housekeeping parts [Eq.~(\ref{eq:MN_decomposition})]. 
    }
    \label{fig:MN_decomp}
\end{figure}

We explain the geometric decomposition of the entropy production rate into excess and housekeeping parts. This geometric decomposition is substantially discussed in terms of optimal transport theory~\cite{Nakazato2021-do}, and geometrically formulated for Langevin systems with uniform temperature~\cite{Dechant2022-gt} and general Markov processes~\cite{yoshimura2023housekeeping}, respectively. This decomposition is mathematically equivalent to the decomposition discussed by Maes and Neto\u{c}n\'{y}~\cite{Maes2014-pk} for systems with a uniform temperature.
Based on a geometric decomposition, the entropy production rate $\sigma_t$ can be decomposed into two non-negative parts (Fig. \ref{fig:MN_decomp}): \begin{align}
\sigma_t=\sigma^\mathrm{hk}_t+\sigma^\mathrm{ex}_t, \label{eq:MN_decomposition}
\end{align}
where $\sigma^\mathrm{ex}_t$ is the excess entropy production rate and $\sigma^\mathrm{hk}_t$ is the housekeeping entropy production rate. The excess entropy production rate $\sigma^\mathrm{ex}_t$ represents the thermodynamic cost caused by changes in the probability distribution $p_t(\bm{x})$ in a non-steady state. In a steady state $\partial p_t(\bm{x})/\partial t=0$, $\sigma^\mathrm{ex}_t=0$. On the other hand, the housekeeping entropy production rate $\sigma^\mathrm{hk}_t$ represents the thermodynamic cost incurred simply to maintain the probability distribution $p_t(\bm{x})$, and occurs even in a steady state (Fig. \ref{fig:MN_decomp}).

The geometric decomposition is based on a decomposition of the local mean velocity $\bm{\nu}_t(\bm{x})$ into excess and housekeeping parts (Fig. \ref{fig:MN_decomp}):
\begin{align}
\bm{\nu}_t(\bm{x})=\bm{\nu}_t^\mathrm{ex}(\bm{x})+\bm{\nu}_t^\mathrm{hk}(\bm{x})\label{eq:MN_decomposition_nu}
\end{align}
Below, we explain the definitions of each term. 

The excess entropy production rate $\sigma^\mathrm{ex}_t$ is defined using a local mean velocity $\bm{\nu}_t^\mathrm{ex}(\bm{x})$, which can be expressed as the gradient of some potential function $\phi_t(\bm{x})$,
\begin{align}
    \bm{\nu}_t^\mathrm{ex}(\bm{x})&=\mathit{D}_t\nabla \phi_t(\bm{x})\label{eq:nu_ex_grad},
\end{align}
and realizes the same time variation of the probability distribution $p_t(\bm{x})$ (Eq. \ref{eq:Fokker--Planck}) as the original local mean velocity $\bm{\nu}_t(\bm{x})$ (Eq. \ref{eq:local_mean_velocity}):
\begin{align}
    \pdv{p_t(\bm{x})}{t}&= -\nabla \cdot \qty[\bm{\nu}_t(\bm{x})p_t(\bm{x})]\nonumber \\
    &=-\nabla \cdot \qty[\bm{\nu}_t^\mathrm{ex} (\bm{x}) p_t(\bm{x})].\label{eq:Fokker--Planck_ex} 
\end{align}
In optimal transport theory, it is known that the velocity field $\bm{\nu}_t^\mathrm{ex}(\bm{x})$ which satisfies this condition is uniquely introduced. The excess entropy production rate $\sigma^\mathrm{ex}_t(\bm{x})$ is defined using this local mean velocity $\bm{\nu}_t^\mathrm{ex}(\bm{x})$ as follows:
\begin{align}
    \sigma^\mathrm{ex}_t =\langle \qty(\bm{\nu}_t^\mathrm{ex})^\top \mathit{D}_t^{-1}\bm{\nu}_t^\mathrm{ex} \rangle_t. \label{eq:EP_ex}
\end{align}

The housekeeping entropy production rate $\sigma^\mathrm{hk}_t$ is defined using the remaining local mean velocity 
\begin{align}
\bm{\nu}_t^\mathrm{hk}(\bm{x})=\bm{\nu}_t(\bm{x})-\bm{\nu}_t^\mathrm{ex} (\bm{x})
\end{align}
as
\begin{align}
\sigma^\mathrm{hk}_t=\langle \qty(\bm{\nu}_t^\mathrm{hk})^\top \mathit{D}_t^{-1}\bm{\nu}_t^\mathrm{hk} \rangle_t.  \label{eq:EP_hk}
\end{align}
From Eq.~(\ref{eq:Fokker--Planck_ex}), we can see that the housekeeping part of the local mean velocity does not change the probability distribution $p_t(\bm{x})$
\begin{align}
    0=-\nabla \cdot \qty[\bm{\nu}_t^\mathrm{hk}(\bm{x})p_t(\bm{x})]. \label{eq:Fokker--Planck_hk}
\end{align}
Hence, the housekeeping entropy production rate $\sigma^\mathrm{hk}_t$ is a thermodynamic cost which does not contribute to the time variation of the probability distribution $p_t(\bm{x})$. We remark that  thermodynamic uncertainty relations~\cite{barato2015thermodynamic}, which provide a fundamental thermodynamic limit of accuracy, have been discussed for this housekeeping entropy production rate and the excess entropy production rate~\cite{dechant2022geometric}.

The sum of the excess entropy production rate $\sigma^\mathrm{ex}_t$ and the housekeeping entropy production rate $\sigma^\mathrm{hk}_t$ equals the total entropy production rate $\sigma_t$ [Eq.~(\ref{eq:MN_decomposition})]. This is because
\begin{align}
\sigma_t&=\langle \qty(\bm{\nu}_t^\mathrm{ex}+\bm{\nu}_t^\mathrm{hk})^\top \mathit{D}_t^{-1}\qty(\bm{\nu}_t^\mathrm{ex}+\bm{\nu}_t^\mathrm{hk}) \rangle_t \nonumber\\
&=\sigma^\mathrm{ex}_t+\sigma^\mathrm{hk}_t+2\langle \qty(\bm{\nu}_t^\mathrm{ex})^\top \mathit{D}_t^{-1}\bm{\nu}_t^\mathrm{hk} \rangle_t
\end{align}
and we find that
\begin{align}
\langle \qty(\bm{\nu}_t^\mathrm{ex})^\top \mathit{D}_t^{-1}\bm{\nu}_t^\mathrm{hk}\rangle_t=&\int \qty[(\nabla \phi_t(\bm{x}_t) )^\top \bm{\nu}_t^\mathrm{hk}(\bm{x}_t) p(\bm{x}_t)] d\bm{x}_t \nonumber\\
=&-\int \phi_t(\bm{x}_t)  \nabla \cdot\qty[\bm{\nu}_t^\mathrm{hk}(\bm{x}_t) p(\bm{x}_t)] d\bm{x}_t \nonumber\\
=& 0, \label{eq:orthogonal}
\end{align}
by using partial integration and Eq.~(\ref{eq:Fokker--Planck_hk}), where we assumed $p_t(\bm{x})\to 0$ when $\|\bm{x}\|\to \infty$.

By definitions [Eqs.~(\ref{eq:EP_ex}) and (\ref{eq:EP_hk})], the excess and housekeeping entropy production rates are non-negative, $\sigma^{\rm ex}_t \geq 0$ and $\sigma^{\rm hk}_t \geq 0$. From the decomposition [Eq.~(\ref{eq:MN_decomposition})], we obtain $\sigma_t \geq \sigma^{\rm ex}_t$ and $\sigma_t \geq \sigma^{\rm hk}_t$. Because the housekeeping entropy production rate $\sigma^\mathrm{hk}_t$ is a thermodynamic cost which does not contribute to $\partial p_t(\bm{x})/\partial t$,  the inequality $\sigma_t \geq \sigma^{\rm ex}_t$ means that the excess entropy production rate $\sigma^\mathrm{ex}_t$ is the minimum entropy production rate for the fixed $\partial p_t(\bm{x})/\partial t$. 

We note that such decomposition of the entropy production rate into the excess part $\sigma^\mathrm{ex}_t$ and that the housekeeping part $\sigma^\mathrm{ex}_t$ is not unique in general~\cite{Dechant2022-gt, dechant2022geometric}, and that geometric decomposition is not equivalent to Hatano-Sasa decomposition in steady-state thermodynamics~\cite{hatano2001steady}. The excess entropy production rate of geometric decomposition can be written using the $L_2$-Wasserstein distance in optimal transport theory~\cite{villani2009optimal} (see Appendix \ref{appendix:excess_wasserstein}).

\section{Housekeeping and excess entropy production rates in Gaussian processes}
\subsection{Time-variation of probability distribution in Gaussian processes}
Here, we discuss a geometric decomposition for Gaussian processes.  We formulate equations that capture the time variation of the probability distribution $p_t (\boldsymbol{x})$ under the assumption of Gaussian processes.
We assume the following linear Langevin equation: \begin{align}
    d\bm{x}_t=\mathit{D}_t[\mathit{A}_t\bm{x}_t+\bm{b}_t]dt+\sqrt{2}\mathit{G}_td\bm{B}_t \label{eq:Langevin_linear}
\end{align}where we set $\bm{f}_t(\bm{x}_t)=\mathit{A}_t\bm{x}_t+\bm{b}_t$ with $d \times d$ matrix $\mathit{A}_t$ and $d$-dimensional vector $\bm{b}_t$ in Eq.~(\ref{eq:Langevin}). The corresponding Fokker-Planck equation is given by 
\begin{align}
\frac{\partial}{\partial t} p_t(\boldsymbol{x}) &= - \nabla \cdot ( \boldsymbol{\nu}_t (\boldsymbol{x}) p_t(\boldsymbol{x})), \nonumber \\
\bm{\nu}_t(\boldsymbol{x}) &= D_t (\mathit{A}_t\bm{x}+\bm{b}_t - \nabla \ln p_t (\boldsymbol{x})). \label{eq:Fokker_linear}
\end{align}
We also assume that the initial probability distribution $p_t (\boldsymbol{x})$ is Gaussian. In the linear Langevin equations [Eq.~(\ref{eq:Langevin_linear})], if the initial probability distribution $p_t(\boldsymbol{x})$ is a Gaussian distribution and the noise term $d\bm{B}_t$ obeys the Gaussian distribution, then the future probability distribution $p_{t+\Delta t} (\boldsymbol{x})$ with $\Delta t >0$ keeps a Gaussian distribution.
Accordingly, it is sufficient to follow the time variation of the mean $\bm{\mu}_t$ and the covariance matrix $\mathit{\Sigma}_t$ of the Gaussian distribution
\begin{align}
p_t(\boldsymbol{x}) = \frac{\exp \left[ - \frac{1}{2}(\boldsymbol{x} - \boldsymbol{\mu}_t)^{\top} \mathit{\Sigma}_t^{-1} (\boldsymbol{x} - \boldsymbol{\mu}_t) \right]}{(2 \pi )^{-\frac{d}{2}} \sqrt{\det \mathit{\Sigma}_t}}.
\end{align}

The time variation of the mean $\bm{\mu}_t$ and the covariance matrix $\mathit{\Sigma}_t$ in the Langevin equation [Eq.~(\ref{eq:Langevin_linear})] are given by
\begin{align}
    \dot{\bm{\mu}}_t=& \frac{1}{dt}\mathbb{E} [d\bm{x}_t  ]\nonumber \\
    =&\mathit{D}_t\mathit{A}_t\bm{\mu}_t+\mathit{D}_t\bm{b}_t, \label{eq:tv_mu}\\
    \dot{\mathit{\Sigma}}_t=&\frac{1}{dt}\left(\mathbb{E} [ (\bm{x}_{t+dt}-\bm{\mu}_{t+dt})(\bm{x}_{t+dt}-\bm{\mu}_{t+dt})^\top ] \right. \nonumber \\
     &-\left.\mathbb{E} [(\bm{x}_t-\bm{\mu}_t)(\bm{x}_t-\bm{\mu}_t)^\top ] \right) \nonumber\\
    =&\mathit{D}_t\mathit{A}_t\mathit{\Sigma}_t+\mathit{\Sigma}_t(\mathit{D}_t\mathit{A}_t)^\top +2\mathit{D}_t, \label{eq:Lyapunov}
\end{align}
where we use $\mathbb{E} [\bm{x}_t] = \boldsymbol{\mu}_t$, $\mathbb{E} [(\bm{x}_{t}-\bm{\mu}_{t})(\bm{x}_{t}-\bm{\mu}_{t})^\top] = \mathit{\Sigma}_t$, $\mathbb{E}[d\boldsymbol{B}_t]= \boldsymbol{0}$ and $\mathbb{E}[d \boldsymbol{B}_{t}d\boldsymbol{B}_{s}^\top]=\delta(t-s) I dt$ and neglect the term $O(dt)$.
Under the assumption of a Gaussian distribution, the time variation of the probability distribution $p_t (\bm{x})$ is equivalent to the following time variations of the mean $\bm{\mu}_t$ and the covariance matrix $\mathit{\Sigma}_t$. Thus, the condition of the steady state, which implies that the probability distribution does not change over time, can be rephrased as $\dot{\bm{\mu}}_t =0$ and $\dot{\mathit{\Sigma}}_t=0$. In the steady state, the equation for the covariance matrix [Eq.~(\ref{eq:Lyapunov})] with  $\dot{\mathit{\Sigma}}_t=0$ is called the Lyapunov equation.

\subsection{Geometric decomposition of entropy production rate in Gaussian processes}
We discuss a geometric decomposition for Eq.~(\ref{eq:Fokker_linear}) under the assumption of a Gaussian process. Under this assumption, the time variation by the excess part of the local mean velocity $\bm{\nu}_t^\mathrm{ex}(\bm{x})$ must provide the same time variation by $\bm{\nu}_t(\bm{x})$. Thus, $\bm{\nu}_t^\mathrm{ex} (\bm{x})$ must be a linear function of $\bm{x}$ because the process is Gaussian. Because $\nabla \ln p_t(\bm{x}) = -\mathit{\Sigma}^{-1}_t (\boldsymbol{x} - \bm{\mu}_t)$ is a linear function of $\bm{x}$,  $\bm{\nu}_t^\mathrm{ex} (\bm{x})$ can be written as
\begin{align}
\bm{\nu}_t^\mathrm{ex}(\bm{x}) &= D_t \nabla \phi_t (\boldsymbol{x}) \nonumber \\
&= \mathit{D}_t\mathit{A}_t^\mathrm{ex}\bm{x}+\mathit{D}_t\bm{b}_t^\mathrm{ex}-\mathit{D}_t\nabla \ln p_t(\bm{x})\label{eq:nu_ex}. 
\end{align}
with some matrix $\mathit{A}_t^\mathrm{ex}$ and vector $\bm{b}_t^\mathrm{ex}$. We remark that $\bm{\nu}_t^\mathrm{ex} (\bm{x})$ cannot be generally written as Eq.~(\ref{eq:nu_ex}) if $p_t(\bm{x})$ is not Gaussian.

Furthermore, $A^\mathrm{ex}_t$ is a symmetric matrix. This is because the excess part of the local mean velocity $\bm{\nu}_t^\mathrm{ex}(\bm{x})$ can be expressed using the gradient of some potential function $\phi_t(\bm{x})$. 
The $(i, j)$-th component of the Hessian of the potential function $\phi$ is 
\begin{align}
    \pdv{x_j}\pdv{x_i}\phi_t (\bm{x}_t)=A^\mathrm{ex}_{ij}+(\mathit{\Sigma}_t^{-1})_{ij}. 
\end{align}
Since the Hessian matrix and the $\mathit{\Sigma}_t^{-1}$ are symmetric, $A_t^\mathrm{ex}$ is also symmetric.

Because the excess part of the local mean velocity $\bm{\nu}_t^\mathrm{ex}(\bm{x})$ realizes the same time variation of the probability distribution as the original local mean velocity $\bm{\nu}_t(\bm{x})$, we obtain the time variation of the mean and the covariance matrix:
\begin{align}
\dot{\bm{\mu}}_t&=\mathit{D}_t\mathit{A}_t\bm{\mu}_t+\mathit{D}_t\bm{b}_t \nonumber \\
&=\mathit{D}_t\mathit{A}_t^\mathrm{ex}\bm{\mu}_t+\mathit{D}_t\bm{b}_t^\mathrm{ex}, \label{eq:tv_mu_ex}\\
\dot{\mathit{\Sigma}}_t&=\mathit{D}_t\mathit{A}_t\mathit{\Sigma}_t+\mathit{\Sigma}_t(\mathit{D}_t\mathit{A}_t)^\top +2\mathit{D}_t \nonumber \\
&=\mathit{D}_t\mathit{A}_t^\mathrm{ex}\mathit{\Sigma}_t+\mathit{\Sigma}_t(\mathit{D}_t\mathit{A}_t^\mathrm{ex})^\top +2\mathit{D}_t. \label{eq:Lyapunov_ex}
\end{align}

In the same way, we can rewrite the housekeeping part of the local mean velocity $\bm{\nu}_t^\mathrm{hk} (\bm{x})$ in Gaussian processes as:
\begin{align}
    \bm{\nu}_t^\mathrm{hk}(\bm{x})=\mathit{D}_t\mathit{A}_t^\mathrm{hk}\bm{x}+\mathit{D}_t\bm{b}_t^\mathrm{hk}, \label{eq:nu_hk_linear}
\end{align}
where \begin{align}
    \mathit{A}_t^\mathrm{hk}&=\mathit{A}_t-\mathit{A}_t^\mathrm{ex},\\
    \bm{b}_t^\mathrm{hk}&=\bm{b}_t-\bm{b}_t^\mathrm{ex}.
\end{align}
From the equations for the time variation of the mean $\bm{\mu}_t$ [Eq.~(\ref{eq:tv_mu_ex})] and the covariance matrix $\mathit{\Sigma}_t$ [Eq.~(\ref{eq:Lyapunov_ex})], we obtain \begin{align}
    \boldsymbol{0}&=\mathit{D}_t\mathit{A}_t^\mathrm{hk}\bm{\mu}_t+\mathit{D}_t\bm{b}_t^\mathrm{hk},\label{eq:tv_mu_hk}\\
     O&=\mathit{D}_t\mathit{A}_t^\mathrm{hk}\mathit{\Sigma}_t+\mathit{\Sigma}_t (\mathit{D}_t\mathit{A}_t^\mathrm{hk})^\top,\label{eq:lyapunov_hk}
\end{align}
where $\boldsymbol{0}$ and $O$ are the zero vector and the zero matrix, respectively. 

By substituting these expressions of $\boldsymbol{\nu}_t (\boldsymbol{x})$,   $\boldsymbol{\nu}^{\rm ex}_t (\boldsymbol{x})$ and $\boldsymbol{\nu}^{\rm hk}_t (\boldsymbol{x})$ in Eqs.~(\ref{eq:Fokker_linear}), (\ref{eq:nu_ex}) and (\ref{eq:nu_hk_linear}) into the definitions of the entropy production rates in Eqs.~(\ref{eq:EP_nu}), (\ref{eq:EP_ex}) and (\ref{eq:EP_hk}), we obtain analytical expressions of the entropy production rates for Gaussian processes,
\begin{align}
    \sigma_t =&(\mathit{A}_t \bm{\mu}_t + \bm{b}_t )^{\top} \mathit{D}_t (\mathit{A}_t \bm{\mu}_t + \bm{b}_t ) \nonumber \\
    &+ \trace \qty[ (\mathit{A}_t + \mathit{\Sigma}_t^{-1} )^{\top} \mathit{D}_t  (\mathit{A}_t + \mathit{\Sigma}_t^{-1} )\mathit{\Sigma}_t], \label{eq:analytical-epr}\\
    \sigma_t^{\rm ex} =&(\mathit{A}^{\rm ex}_t \bm{\mu}_t + \bm{b}^{\rm ex}_t )^{\top} \mathit{D}_t (\mathit{A}^{\rm ex}_t \bm{\mu}_t + \bm{b}^{\rm ex}_t ) \nonumber \\
    &+ \trace \qty[ (\mathit{A}^{\rm ex}_t + \mathit{\Sigma}_t^{-1} )^{\top} \mathit{D}_t  (\mathit{A}^{\rm ex}_t + \mathit{\Sigma}_t^{-1} )\mathit{\Sigma}_t],\label{eq:analytical-excess-epr} \\
    \sigma^\mathrm{hk}_t =& \trace \qty[ (\mathit{A}^{\rm hk}_t)^{\top} \mathit{D}_t  \mathit{A}^{\rm hk}_t \mathit{\Sigma}_t].\label{eq:analytical-housekeeping-epr}
\end{align}
We explain geometric interpretations of this decomposition of the entropy production rate using the Hilbert--Schmidt inner product in Appendix \ref{appendix:HS-innerproduct}.

\section{Oscillatory mode decomposition of virtual dynamics given by $\bm{\nu}_t^\mathrm{hk}$}

In preparation for the derivation of the main results in the next section, we show that the housekeeping part of the local mean velocity $\bm{\nu}_t^\mathrm{hk} (\bm{x})$ in Eq.~(\ref{eq:nu_hk_linear}) reflects oscillatory dynamics. 
To observe this, we consider the following virtual deterministic process: 
\begin{align}
d\bm{x}_s=\qty[\mathit{D}_t\mathit{A}_t^\mathrm{hk}\bm{x}_s+\mathit{D}_t\bm{b}_t^\mathrm{hk}]ds =\bm{\nu}_t^\mathrm{hk}(\bm{x}_s) ds\label{eq:Langevin_hk}. 
\end{align}
where the housekeeping part of the local mean velocity $\bm{\nu}_t^\mathrm{hk} (\bm{x})$ is introduced by the original process [Eq.~(\ref{eq:Langevin_linear})]. Here, $s$ stands for the time of the virtual dynamics, whereas $t$ stands for the time of the original Langevin dynamics [Eq.~(\ref{eq:Langevin_linear})].
During the virtual deterministic processes, the function $\bm{\nu}_t^\mathrm{hk} (\bm{x})$ is fixed with respect to changes in $s$.
We here consider the continuity equation $\partial q_s(\boldsymbol{x})/\partial s = -\nabla \cdot(\bm{\nu}_t^\mathrm{hk}(\bm{x}) q_s(\bm{x}))$ for the time variation of the probability distribution $q_s(\bm{x})$ for this deterministic virtual process. If $q_s(\bm{x})$ is given by the same distribution in the original dynamics $p_t(\bm{x})$, then $\left. \partial q_s(\bm{x})/\partial s \right|_{q_s(\bm{x})=p_t(\bm{x})}=
-\nabla \cdot(\bm{\nu}_t^\mathrm{hk}(\bm{x}) p_t(\bm{x})) =0$. This fact implies that  $p_t(\bm{x})$ is the invariant measure of this virtual deterministic process [Eq.~(\ref{eq:Langevin_hk})]. 

We next discuss the analytical solution for the future state $\bm{x}_{s+\Delta s}$ with $\Delta s>0$ in the virtual deterministic process [Eq.~(\ref{eq:Langevin_hk})]. From Eq.~(\ref{eq:tv_mu_hk}) and $d\bm{\mu}_t/ds= \boldsymbol{0}$, we obtain  $d[\bm{x}_s-\bm{\mu}_t]/ds =\mathit{D}_t\mathit{A}_t^\mathrm{hk} \qty[\bm{x}_s-\bm{\mu}_t]$, and thus
\begin{align}
    \bm{x}_{s+\Delta s}&=\bm{\mu}_t+e^{\mathit{D}_t A_t^\mathrm{hk}\Delta s}\qty(\bm{x}_s-\bm{\mu}_t)\nonumber\\
    &=\bm{\mu}_t+\sum_k e^{\lambda_k \Delta s}\mathsf{F}_k\qty(\bm{x}_s-\bm{\mu}_t)\nonumber\\
    &=\bm{\mu}_t+\sum_k e^{2\pi\chi_k \mathrm{i} \Delta s}\mathsf{F}_k\qty(\bm{x}_s-\bm{\mu}_t). \label{eq:oscillation}
\end{align}
Here, $\lambda_k$ is the $k$-th eigenvalue of $\mathit{D}_tA_t^\mathrm{hk}$ and  
$\chi_k$ is defined as \begin{align}
    \chi_k = \lambda_k/(2\pi \mathrm{i}) \label{eq:def_chi},
\end{align} where $\mathrm{i}$ stands for  the imaginary unit. As proved later, $\lambda_k$ is purely imaginary, and hence, $\chi_k$ is a real number. 
$\mathsf{F}_k$ is the projection matrix that provides the spectral decomposition of $\mathit{D}_tA_t^\mathrm{hk}$, 
\begin{align}
\mathit{D}_tA_t^\mathrm{hk} = \sum_k \lambda_k \mathsf{F}_k. \label{eq:eigenvalue_decomposition}
\end{align}
This projection matrix $\mathsf{F}_k$ is introduced using
the eigenvalue decomposition of $\mathit{D}_tA_t^\mathrm{hk}$ as $\mathit{D}_tA_t^\mathrm{hk}=\mathsf{P}\Lambda \mathsf{P}^{-1}$, where $\mathsf{P}$ is a matrix of eigenvectors, and $\Lambda$ is a diagonal matrix whose $k$-th element is the $k$-th eigenvalue $\lambda_k$.
Let $\bm{e}_k$ be the $k$-th basic unit vector, whose $k$-th element is $1$ and all other elements are $0$. This projection matrix $\mathsf{F}_k$ is explicitly defined as
\begin{align}
    \mathsf{F}_k:=\mathsf{P}\bm{e}_k\bm{e}_k^\top \mathsf{P}^{-1} ,\label{eq:F_hk_def}
\end{align} 
and the eigenvalue decomposition was rewritten as Eq.~(\ref{eq:eigenvalue_decomposition}). 

The time evolution of the virtual dynamics [Eq.~(\ref{eq:oscillation})] is rewritten as the time evolution of the modes. Since $\sum_k \bm{e}_k\bm{e}_k^\top =\mathit{I}$, the projection matrix satisfies $\sum_k \mathsf{F}_k=I$. Thus, $\bm{x}_s - \bm{\mu}_t$ can be decomposed as the sum of the modes $\mathsf{F}_k\qty(\bm{x}_s-\bm{\mu}_t)$,
\begin{align}
    \bm{x}_s-\bm{\mu}_t=\sum_k \mathsf{F}_k\qty(\bm{x}_s-\bm{\mu}_t),
\end{align}
and thus Eq.~(\ref{eq:oscillation}) reads the time evolution of the modes
\begin{align}
    \sum_{k} \mathsf{F}_k (\bm{x}_{s+\Delta s} - \bm{\mu}_t) &=\sum_k e^{2\pi \chi_k \mathrm{i}\Delta s}\mathsf{F}_k\qty(\bm{x}_s-\bm{\mu}_t), \label{eq:oscillation2}
\end{align}
where we used $\sum_{k} \mathsf{F}_k (\bm{x}_{s+\Delta s} - \bm{\mu}_t) =\bm{x}_{s+\Delta s} - \bm{\mu}_t $.
We can also show that each mode evolves independently as follows,
\begin{align}
    \mathsf{F}_k\qty(\bm{x}_{s+\Delta s}-\bm{\mu}_t)=e^{2\pi \chi_k \mathrm{i}\Delta s} \mathsf{F}_k\qty(\bm{x}_s-\bm{\mu}_t),
    \label{eq:oscillation3}
\end{align}
where we multiplied both sides of Eq.~(\ref{eq:oscillation2}) by $\mathsf{F}_k$ from the left, and used a mathematical property of the projection matrix $\mathsf{F}_k \mathsf{F}_j = \delta_{k j} \mathsf{F}_k$.

Because the eigenvalues of $\mathit{D}_t \mathit{A}_t^\mathrm{hk}$ are $0$ or purely imaginary, the mode $\mathsf{F}_k\qty(\bm{x}_s-\bm{\mu}_t)$ can be regarded as the oscillatory mode. To show this fact, we use Eq.~(\ref{eq:lyapunov_hk}), and $A_t^\mathrm{hk}$ can be written as
\begin{align}
\mathit{D}_t\mathit{A}_t^\mathrm{hk}=\mathit{Q}\mathit{\Sigma}_t^{-1}
\end{align}
with an anti-symmetric matrix $\mathit{Q}=-\mathit{Q}^\top$. 
This is because $D_tA_t^\mathrm{hk}\mathit{\Sigma}_t$ is an anti-symmetric matrix from 
Eq. (\ref{eq:lyapunov_hk}). 
By denoting the eigenvalues of a matrix $X$ as $\mathrm{eig}(X)$, the eigenvalues $\mathit{D}_t\mathit{A}_t^\mathrm{hk}$ can be written as
\begin{align}
\mathrm{eig}\qty(\mathit{D}_t\mathit{A}_t^\mathrm{hk})=\mathrm{eig}\qty(\mathit{Q}\mathit{\Sigma}_t^{-1})=\mathrm{eig}\qty( (\sqrt{\mathit{\Sigma}_t})^{-1}\mathit{Q} (\sqrt{\mathit{\Sigma}_t})^{-1} ), \label{eq:eig_hk}
\end{align}
where we use the fact that $\mathrm{eig}(XY)=\mathrm{eig}(YX)$ for any pair of matrices $X$ and $Y$. 
Since $ (\sqrt{\mathit{\Sigma}_t})^{-1}\mathit{Q} (\sqrt{\mathit{\Sigma}_t})^{-1} $ is an anti-symmetric matrix, and the eigenvalues of an anti-symmetric matrix are $0$ or purely imaginary, the eigenvalues of $\mathit{D}_t A_t^\mathrm{hk}$ are also $0$ or purely imaginary. This implies that $\chi_k=\lambda_k/(2\pi\mathrm{i})$ is a real number. 
Hence, Eq.~(\ref{eq:oscillation3}) means that the $k$-th mode $\mathsf{F}_k\qty(\bm{x}_s-\bm{\mu}_t)$ temporally oscillates with the frequency $\chi_k$, and Eq.~(\ref{eq:oscillation}) means that the time evolution of the virtual dynamics is given by the sum of the temporal oscillatory modes. Here we still consider the invariant measure $p_t(\boldsymbol{x})$, and these temporal oscillations do not contribute to the time evolution of $p_t(\boldsymbol{x})$.

Further, we discuss the intensity of the oscillatory mode in the original process [Eq.~(\ref{eq:Langevin_linear})]. By replacing $\bm{x}_s$ with $\bm{x}_t$ in $\mathsf{F}_k\qty(\bm{x}_s-\bm{\mu}_t)$, we introduce the $k$-th oscillatory mode $\mathsf{F}_k\qty(\bm{x}_t-\bm{\mu}_t)$ in the original process. The intensity of the $k$-th mode is introduced as the expected value of the square of the Hilbert-Schmidt norm:
\begin{align}
\langle \|\mathsf{F}_k(\bm{x}_t-\bm{\mu}_t) \|^2_{\rm HS} \rangle_t &=  \langle\trace [{ [\mathsf{F}_k (\bm{x}_t-\bm{\mu}_t)]^* \mathsf{F}_k(\bm{x}_t-\bm{\mu}_t) }]\rangle_t \nonumber \\
 &= \trace ( {\mathsf{F}_k\mathit{\Sigma}_t \mathsf{F}_k^*}),
\end{align}
 where the symbol $^*$ stands for the conjugate transpose of the matrix and the Hilbert-Schmidt norm for the complex matrix $Y$ is defined as $\| Y \|_{\rm HS} =\sqrt{\trace(Y^* Y)}$. We note that this value $\trace ( {\mathsf{F}_k\mathit{\Sigma}_t \mathsf{F}_k^*})$ is also regarded as the intensity of the oscillatory mode in the virtual deterministic process,
 \begin{align}
\langle \|\mathsf{F}_k(\bm{x}_s-\bm{\mu}_t) \|^2_{\rm HS} \rangle_{\rm inv}  &= \langle \|\mathsf{F}_k(\bm{x}_t-\bm{\mu}_t) \|^2_{\rm HS} \rangle_t\nonumber \\
&= \trace ( {\mathsf{F}_k\mathit{\Sigma}_t \mathsf{F}_k^*}),
\end{align}
where $\langle \cdots \rangle_{\rm inv}$ is the expectation with respect to the invariant measure $p_t$, defined as $\langle \cdots \rangle_{\rm inv} = \int d \bm{x}_s p_t(\bm{x}_s) \cdots$.

We also introduce the $i$-th intensity of the $k$-th oscillatory mode by considering each diagonal element of the matrix $\mathsf{F}_k\mathit{\Sigma}_t \mathsf{F}_k^*$. The $i$-th intensity of the $k$-th oscillatory mode is computed as the expected value of the square of the norm:
\begin{align}
\langle \|(\mathsf{F}_k\qty(\bm{x}_t-\bm{\mu}_t))_i\|^2 \rangle_t &= \langle \|(\mathsf{F}_k\qty(\bm{x}_s-\bm{\mu}_t))_i\|^2 \rangle_{\rm inv} \nonumber \\
&=(\mathsf{F}_k\mathit{\Sigma}_t \mathsf{F}_k^*)_{ii}. \label{eq:intensity}
\end{align}

\section{Mode decomposition of the housekeeping entropy production rate via oscillatory modes}
\subsection{Main result}
As the main result of this paper, we derived  a decomposition of the housekeeping entropy production rate into independent positive contributions for each oscillatory mode:
\begin{align}
    \sigma^\mathrm{hk}_t &= (2\pi)^2\sum_k \chi_k^2\tr(\mathit{G}_t^{-1}\mathsf{F}_k\mathit{\Sigma}_t \mathsf{F}_k^*(\mathit{G}_t^{-1})^{\top}) \nonumber \\
    &=(2\pi)^2\sum_k \chi_k^2\| \mathit{G}_t^{-1}\mathsf{F}_k \sqrt{\mathit{\Sigma}_t} \|^2_{\rm HS}
    \label{eq:EP_hk_decomposition},
\end{align}
where we assume that the eigenvalues of $D_t\mathit{A}_t^\mathrm{hk}$ are not degenerate.
Because the Hilbert-Schmidt norm $\| \mathit{G}_t^{-1}\mathsf{F}_k \sqrt{\mathit{\Sigma}_t} \|_{\rm HS}$ is non-negative, we can see that the housekeeping entropy production rate can be decomposed into independent positive contributions from each oscillatory mode.

When the noise coefficient $\mathit{G}_t$ is a diagonal matrix, the meaning of the mode decomposition is more evident:
\begin{align}
\sigma^\mathrm{hk}_t = (2\pi)^2\sum_k\chi_k^2 \sum_i (D_t^{-1})_{ii}(\mathsf{F}_k\mathit{\Sigma}_t \mathsf{F}_k^*)_{ii}\label{eq:EP_hk_decomposition_uniform}.
\end{align}
Recall that the diagonal elements of $\mathsf{F}_k\mathit{\Sigma}_t \mathsf{F}_k^*$ correspond to the intensity of the $k$-th oscillatory mode [(Eq. \ref{eq:intensity})]. \begin{align}
    (D_t^{-1})_{ii}(\mathsf{F}_k\mathit{\Sigma}_t \mathsf{F}_k^*)_{ii} \label{eq:normalized_intensity}
\end{align} is the intensity normalized by the diffusion matrix $D_t$. As discussed, $\chi_k$ corresponds to the frequency of oscillation. Hence, we can see that the contributions of each oscillatory mode to the housekeeping entropy production rate can be written by contributions of the normalized intensity and the square of the frequency of the oscillation.
In other words, oscillatory modes with higher frequency and normalized intensity contribute more to the housekeeping entropy production rate. We note that these oscillations can be regarded as temporal oscillations in the virtual dynamics, and that these temporal oscillations do not contribute to the transient change in probability.  
We also note that the normalized intensity is attributed to $\mathit{\Sigma}_t$, which determines the spatial shape of the probability distribution $p_t(\boldsymbol{x})$.

Since the normalized intensity is attributed to each element of $\bm{x}_t$, we can also examine the contribution of each element of $\bm{x}$ within the oscillatory modes to the housekeeping entropy production rate.

\subsection{Derivation of the main result}
From the analytical expression of the housekeeping entropy production rate [Eq.~(\ref{eq:analytical-housekeeping-epr})] and the eigenvalue decomposition of $\mathit{D}_t\mathit{A}_t^\mathrm{hk}$ [Eq.~(\ref{eq:eigenvalue_decomposition})], we obtain
\begin{align}
    \sigma^\mathrm{hk}_t&=\trace\qty((\mathit{A}_t^\mathrm{hk})^{\top}
    \mathit{D}_t \mathit{A}_t^\mathrm{hk} \mathit{\Sigma}_t)\nonumber\\
    &=\trace\qty(\mathit{G}_t^{-1}\qty(\mathit{D}_t\mathit{A}_t^\mathrm{hk})\mathit{\Sigma}_t\qty(\mathit{D}_t\mathit{A}_t^{\mathrm{hk}})^*(\mathit{G}_t^{-1})^{\top})\nonumber\\
    &=\sum_{kl}\lambda_k\overline{\lambda_l}\trace \qty(\mathit{G}_t^{-1}\mathsf{F}_k\mathit{\Sigma}_t\mathsf{F}_l^* (\mathit{G}_t^{-1})^{\top} ), \label{eq:sigma_hk_prepare}
\end{align}
where $\overline{\lambda_l}$ stands for the complex conjugate of $\lambda_l$.
From Eq.~(\ref{eq:lyapunov_hk}) and the eigenvalue decomposition of $\mathit{D}_t\mathit{A}_t^\mathrm{hk}$ [Eq.~(\ref{eq:eigenvalue_decomposition})], we obtain
\begin{align}
    0&=\sum_m \qty[\lambda_m \mathsf{F}_m\mathit{\Sigma}_t+\overline{\lambda_m}\mathit{\Sigma}_t \mathsf{F}_m^*]\nonumber\\
    &=\sum_m \lambda_m \qty[ \mathsf{F}_m\mathit{\Sigma}_t-\mathit{\Sigma}_t \mathsf{F}_m^* ],
\end{align}
where we use the fact that the eigenvalue $\lambda_m$ is purely imaginary. 
By multiplying $\mathsf{F}_k$ from the left side and $\mathsf{F}_l^*$ from the right side and using the property of the projection matrix $\mathsf{F}_k \mathsf{F}_j = \delta_{kj} \mathsf{F}_k$, we obtain
\begin{align}
    0&=\lambda_k \mathsf{F}_k\mathit{\Sigma}_t \mathsf{F}_l^*-\lambda_l\mathsf{F}_k\mathit{\Sigma}_t \mathsf{F}_l^*\nonumber\\
    &=(\lambda_k-\lambda_l) \mathsf{F}_k\mathit{\Sigma}_t \mathsf{F}_l^*. \label{eq:orthogonal_hk}
\end{align}
This equation means that $\mathsf{F}_k\mathit{\Sigma}_t \mathsf{F}_l^*=0$ when $\lambda_k\neq\lambda_l$. From
the assumption that the eigenvalues of $\mathit{D}_t\mathit{A}_t^\mathrm{hk}$ are not degenerate, we obtain  $\mathsf{F}_k\mathit{\Sigma}_t \mathsf{F}_l^*= \delta_{kl} \mathsf{F}_k\mathit{\Sigma}_t \mathsf{F}_k^*$ because $\lambda_k\neq\lambda_l$ when $k \neq l$. Thus, Eq.~(\ref{eq:sigma_hk_prepare}) can be rewritten as
\begin{align}
    \sigma^\mathrm{hk}_t
    &=\sum_{k} |\lambda_k|^2 \trace \qty(\mathit{G}_t^{-1}\mathsf{F}_k\mathit{\Sigma}_t\mathsf{F}_k ^* (\mathit{G}_t^{-1})^{\top} )\nonumber\\
    &=(2\pi)^2\sum_{k} \chi_k^2 \trace \qty(\mathit{G}_t^{-1}\mathsf{F}_k\mathit{\Sigma}_t\mathsf{F}_k ^* (\mathit{G}_t^{-1})^{\top} ),
\end{align}
which is our main result [Eq.~(\ref{eq:EP_hk_decomposition})], where $|\cdot|$ stands for the absolute value. 

We remark that our result can be generalized without the assumption of non-degeneracy by using the fact that $\mathsf{F}_k\mathit{\Sigma}_t \mathsf{F}_l^*=0$ when $\lambda_k\neq\lambda_l$. If $\mathit{D}_t\mathit{A}_t^\mathrm{hk}$ is degenerate, the housekeeping entropy production rate [Eq.~(\ref{eq:sigma_hk_prepare})] is calculated as
\begin{align}
    \sigma^\mathrm{hk}_t=\sum_{i \in \mathcal{M}_{\rm I}}  \sum_{k \in \mathcal{D}(i)} \sum_{l \in \mathcal{D}(i)}|\lambda_i|^2 \trace \qty(\mathit{G}_t^{-1}  \mathsf{F}_k \mathit{\Sigma}_t  \mathsf{F}_l^* (\mathit{G}_t^{-1})^{\top} ),
\end{align}
where $\mathcal{M}_{\rm I}$ is the set of the independent modes such that $\lambda_i \neq \lambda_{i'}$ for any $i \in \mathcal{M}_{\rm I}$, $i'\in \mathcal{M}_{\rm I}$, and 
$\mathcal{D}(i)$ is the set of the $i$-th degenerate modes such that $\lambda_k = \lambda_{k'}$ for any $k \in \mathcal{D}(i)$, $k'\in \mathcal{D}(i)$. Thus, we obtain a similar expression
\begin{align}
    \sigma^\mathrm{hk}_t=(2\pi)^2\sum_{i \in \mathcal{M}_{\rm I}} \chi_i^2 \trace \qty(\mathit{G}_t^{-1}   \tilde{\mathsf{F}}_i \mathit{\Sigma}_t \tilde{\mathsf{F}}_i^* (\mathit{G}_t^{-1})^{\top} ),
\end{align}
where $\tilde{\mathsf{F}}_i = \sum_{k \in \mathcal{D}(i)} \mathsf{F}_k$.

\subsection{Interpretation of our decomposition}
In this subsection, we provide an interpretation of our decomposition [Eq. (\ref{eq:EP_hk_decomposition_uniform})] and define several quantities in preparation for the following sections. 

To clarify that the proposed decomposition can be simultaneously decomposed in terms of both oscillatory mode and dimension when the noise matrix $D_t$ is a diagonal matrix, we introduce the following quantities:
\begin{align}
    \sigma_t^\mathrm{hk}&=\sum_i \sum_k \sigma_t^{\mathrm{hk}(k,i)},\label{eq:decomp_oscillation}\\
    \sigma_t^{\mathrm{hk}(k,i)}&=(2\pi\chi_k)^2J_k^{(i)},\label{eq:def_sigma_hk_k_i}\\
    J_k^{(i)}&=(D_t^{-1})_{ii}(\mathsf{F}_k\mathit{\Sigma}_t\mathsf{F}_k^*)_{ii}. \label{eq:def_J_k_i}
\end{align}
First, in Eq. (\ref{eq:decomp_oscillation}), the housekeeping entropy production rate is decomposed into the sum of the contributions of the $i$-th dimension of the $k$-th oscillatory mode, $\sigma_t^{\mathrm{hk}(k,i)}$. Then, in Eq. (\ref{eq:def_sigma_hk_k_i}), $\sigma_t^{{\mathrm{hk}},(k,i)}$ is expressed using the product of the square of the frequency of the $k$-th oscillatory mode, $\chi_k$, and  the normalized intensity of the $i$-th dimension of the $k$-th oscillatory mode $J_k^{(i)}$. Finally, in Eq. (\ref{eq:def_J_k_i}),  $J_k^{(i)}$ is the intensity of the $i$-th dimension of the $k$-th oscillatory mode $(\mathsf{F}_k\mathit{\Sigma}_t\mathsf{F}_k^*)_{ii}$ divided by the $i$-th diagonal element of the diffusion matrix $(D_t)_{ii}$.

Additionally, we have prepared some quantities for next section. 
\begin{align}
    \sigma_t^{\mathrm{hk}(k)}&=\sum_i \sigma_t^{\mathrm{hk}(k,i)}=(2\pi\chi_k)^2J_k,\label{eq:def_sigma_k}\\
    J_k&=\sum_i J_k^{(i)}, \label{eq:def_J_k}
\end{align}
where $\sigma_t^{\mathrm{hk}(k)}$ is a contribution of the $k$-th oscillatory mode and defined as the sum of $\sigma_t^{\mathrm{hk}(k,i)}$ over dimension $i$ [Eq. (\ref{eq:def_sigma_k})]. Note that the sum of $\sigma_t^{\mathrm{hk}(k)}$ over oscillatory modes $k$ recovers the total housekeeping entropy production rate: $\sigma_t^\mathrm{hk}=\sum_k\sigma_t^{\mathrm{hk}(k)}$ [Eq. (\ref{eq:decomp_oscillation})]. Here, $J_k$ is the sum of the normalized intensity $J_k^{(i)}$ over the dimension [Eq. (\ref{eq:def_J_k})], and $\sigma_t^{\mathrm{hk}(k)}$ is written using the product of the square of the frequency $\chi_k$ and $J_k$ [Eq. (\ref{eq:def_sigma_k})].

\section{Examples}
\begin{figure}[t]
\centering
\includegraphics[width=\linewidth]{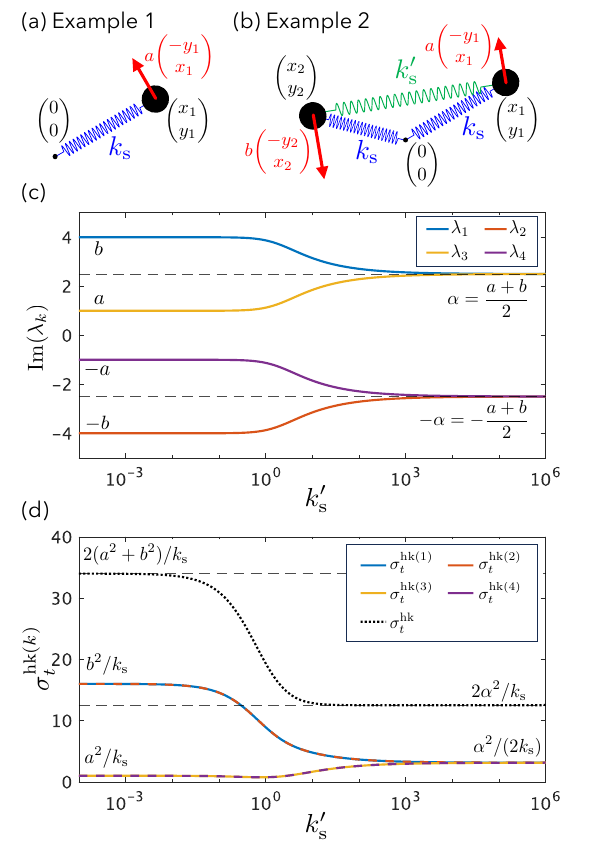}
\caption{
Examples of our decomposition [Eq. (\ref{eq:decomp_oscillation})]. 
(a) An illustration of Example 1. 
(b) An illustration of Example 2. 
(c) Numerical calculation of the eigenvalues $\lambda_k$ for various $k'_{\rm s}$ values in Example 2. 
(d) Numerical calculation of the decomposition of the housekeeping entropy production rate $\sigma_t^{\mathrm{hk}(k)}$ for various $k'_{\rm s}$ values in the example 2. 
In (c) and (d), parameters are set to $a=1, b=4, k_{\rm s}=1$. 
}
\label{fig:fig_toy_example}
\end{figure}

To gain insight, we illustrate our decomposition with two examples in this section. In Example 1, to understand our decomposition, we prepare an analytically tractable system in which a 2D particle receives a rotational force (Fig. \ref{fig:fig_toy_example}a). This is one of the easiest systems to use to gain insight into our decomposition, because applying our decomposition to this system yields only a conjugate pair of oscillatory modes, and we can confirm the relationship between the oscillation frequency and the housekeeping entropy production rate. This decomposition also serves as an example of the decomposition of the entropy production rate into excess and housekeeping parts, since the system is in a non-steady state. The steady-state solution of this example also serves as preparation for the interpretation of Example 2. In Example 2, we consider 2D spring-linked particles which receive rotational forces (Fig. \ref{fig:fig_toy_example}b). This example is helpful for understanding the case in which there are two conjugate pairs of the oscillatory modes with different frequencies. 

\subsection{Example 1: Analytical solution for a 2D particle with rotational force in a non-steady state}
In the first example, we consider a two-dimensional particle whose position is represented as $\bm{x}_t=(x_1, y_1)^\top$. The particle receives a rotational force $(-ay_1,ax_1)^\top$ and is connected to a spring with a spring constant $k_{\rm s}$, which is fixed at the origin as shown in Fig. \ref{fig:fig_toy_example}a:
\begin{align}
    d\mqty(x_1\\y_1)&=\mqty(
      -k_{\rm s}x_1-ay_1 \\
      -k_{\rm s}y_1+ax_1 
    )dt+\sqrt{2}G_td\bm{B}_t.
\end{align}
We set 
\begin{align}
G_t=\mqty(\sqrt{T}&0\\0&\sqrt{T}), \hspace{5mm} D_t=\mqty(T&0\\0&T),
\end{align}
where a positive scalar $T$ is the temperature of the bath. This equation corresponds to 
\begin{align}
    D_tA_t=\mqty(-k_{\rm s}&-a\\a&-k_{\rm s}), \hspace{5mm} D_t\bm{b}_t=\mqty(0\\0)
\end{align}
in the Langevin equation (\ref{eq:Langevin_linear}). The mean values and the variance-covariance matrix are written as
\begin{align}
    \bm{\mu}_t=\mqty(\mu_1\\\mu_2),
    \hspace{5mm}
    \mathit{\Sigma}_t = 
    \mqty(\mathit{\Sigma}_{11} & \mathit{\Sigma}_{12}\\
        \mathit{\Sigma}_{12} & \mathit{\Sigma}_{22}).
\end{align}
We assume here that the variance-covariance matrix is regular, and thus $\det \mathit{\Sigma}_t \neq 0$. We note that the system is not necessarily in the steady state. In this subsection, we first derive the decomposition in the non-steady state, and then derive it in the steady state. 

The excess and housekeeping parts of $D_tA_t$ and $D_t\bm{b}_t$ are solved as the solutions of Eqs. (\ref{eq:tv_mu_ex}), (\ref{eq:Lyapunov_ex}), (\ref{eq:tv_mu_hk}) and (\ref{eq:lyapunov_hk}), under the condition that $A_t^\mathrm{ex}$ is a symmetric matrix: 
\begin{align}
    D_tA_t^\mathrm{ex}&=\mqty(
        -k_{\rm s} - \frac{2a\mathit{\Sigma}_{12}}{\tr\mathit{\Sigma}_t} & a\frac{\mathit{\Sigma}_{11}-\mathit{\Sigma}_{22}}{\tr\mathit{\Sigma}_t}\\
        a\frac{\mathit{\Sigma}_{11}-\mathit{\Sigma}_{22}}{\tr\mathit{\Sigma}_t} & -k_{\rm s} +\frac{2a\mathit{\Sigma}_{12}}{\tr\mathit{\Sigma}_t}
    )\\
    D_t\bm{b}_t^\mathrm{ex}&=\frac{2a}{\tr \mathit{\Sigma}_t}
    \mqty(    
        - \mathit{\Sigma}_{12} \mu_1 + \mathit{\Sigma}_{11} \mu_2\\
        -\mathit{\Sigma}_{22}\mu_1 + \mathit{\Sigma}_{12} \mu_2
    )
\end{align}
and
\begin{align}
    D_tA_t^\mathrm{hk}&=-\frac{2a}{\tr \mathit{\Sigma}_t}\mqty(
          -\mathit{\Sigma}_{12}&\mathit{\Sigma}_{11}\\
           -\mathit{\Sigma}_{22}&\mathit{\Sigma}_{12}
        )\\
    D_t\bm{b}_t^\mathrm{hk}&=-\frac{2a}{\tr \mathit{\Sigma}_t}
    \mqty(    
        - \mathit{\Sigma}_{12} \mu_1 + \mathit{\Sigma}_{11} \mu_2\\
        -\mathit{\Sigma}_{22}\mu_1 + \mathit{\Sigma}_{12} \mu_2
    ).
\end{align}
The eigenvalues of $D_tA_t^\mathrm{hk}$ are
\begin{align}
    \lambda_1=\frac{2a\sqrt{\det \mathit{\Sigma}_t}}{\tr \mathit{\Sigma}_t}\mathrm{i}, \hspace{5mm} \lambda_2=-\frac{2a\sqrt{\det \mathit{\Sigma}_t}}{\tr \mathit{\Sigma}_t}\mathrm{i}.
\end{align}
Thus, the eigenvalues are proportional to the intensity of the rotational force $a$.

Our decomposition of the housekeeping entropy production rate is computed from the spectral decomposition of $D_tA_t^\mathrm{hk}$. The corresponding projection matrices for the spectral decomposition of $D_tA_t^\mathrm{hk}$ are respectively
\begin{align}
    \mathsf{F}_1= \frac{1}{2\mathrm{i} \sqrt{\det \mathit{\Sigma}_t }}\mqty(
        \mathit{\Sigma}_{12} + \sqrt{\det \mathit{\Sigma}_t} \mathrm{i}& - \mathit{\Sigma}_{11}\\
        \mathit{\Sigma}_{22} & -\mathit{\Sigma}_{12}+ \sqrt{\det \mathit{\Sigma}_t} \mathrm{i}
    )\\    
    \mathsf{F}_2= \frac{1}{2\mathrm{i} \sqrt{\det \mathit{\Sigma}_t }}\mqty(
        -\mathit{\Sigma}_{12} + \sqrt{\det \mathit{\Sigma}_t} \mathrm{i}& \mathit{\Sigma}_{11}\\
        -\mathit{\Sigma}_{22} & \mathit{\Sigma}_{12}+ \sqrt{\det \mathit{\Sigma}_t} \mathrm{i}
    ),
\end{align}
because $\mathsf{F}_k$ is written using the eigenvector matrix $\mathsf{P}$ as $\mathsf{F}_k=\mathsf{P}\bm{e}_k\bm{e}^\top \mathsf{P}^{-1}$, where \begin{align}
    \mathsf{P}=\mqty(
        \mathit{\Sigma}_{11}& \mathit{\Sigma}_{11} \\
        \mathit{\Sigma}_{12} - \sqrt{\det \mathit{\Sigma}_t} \mathrm{i}&\mathit{\Sigma}_{12} + \sqrt{\det \mathit{\Sigma}_t}\mathrm{i}
    ).
\end{align}
We can check that $D_tA_t^\mathrm{hk} = \lambda_1 \mathsf{F}_1 + \lambda_2 \mathsf{F}_2$ is satisfied. The normalized intensities $J_k = \sum_i (D_t^{-1})_{ii} (\mathsf{F}_k \mathit{\Sigma}_t \mathsf{F}_k^*)_{ii}$ in Eq. (\ref{eq:def_J_k}), which are used for our decomposition, are 
\begin{align}
    J_1=\frac{1}{2T}\tr\mathit{\Sigma}_t, \hspace{5mm} J_2=\frac{1}{2T}\tr\mathit{\Sigma}_t.
\end{align}
Using $\lambda_k$ and $J_k$, we obtain the contribution of each mode to the housekeeping entropy production rate as $\sigma_t^{\mathrm{hk}(k)}=|\lambda_k|^2J_k$. These are
\begin{align}
    \sigma_t^{\mathrm{hk}(1)}=\frac{2a^2}{T}\frac{\det \mathit{\Sigma}_t}{\tr \mathit{\Sigma}_t}, \hspace{5mm} \sigma_t^{\mathrm{hk}(2)}=\frac{2a^2}{T} \frac{\det \mathit{\Sigma}_t}{ \tr \mathit{\Sigma}_t}. 
\end{align}
The sum of these contributions $\sigma_t^\mathrm{hk}= \sigma_t^{\mathrm{hk}(1)} + \sigma_t^{\mathrm{hk}(2)}$ equals the housekeeping entropy production rate $\sigma_t^\mathrm{hk}$ calculated from Eq. (\ref{eq:analytical-housekeeping-epr}),
\begin{align}
\sigma_t^\mathrm{hk}=\tr[(A_t^\mathrm{hk})^\top D_t A_t^\mathrm{hk}\mathit{\Sigma}_t]=\frac{4a^2}{T}\frac{\det \mathit{\Sigma}_t}{\tr \mathit{\Sigma}_t}.
\end{align}
This shows that our decomposition works, meaning that the sum of the contributions of each mode, calculated from the square of the frequency and the normalized intensity, is equal to the housekeeping entropy production rate. We note that the housekeeping entropy production rate does not explicitly depend on the spring constant $k_{\rm s}$, and it is proportional to $a^2$.

Next, we derive the analytical solution for the steady state. In a steady state, the mean $\bm{\mu}_t$ and the covariance matrix $\mathit{\Sigma}_t$ are solved from (\ref{eq:tv_mu_ex}) and (\ref{eq:Lyapunov_ex}) as follows
\begin{align}
    \bm{\mu}_t=\mqty(0\\0), \hspace{5mm} \mathit{\Sigma}_t=\mqty(\frac{T}{k_{\rm s}}&0\\0&\frac{T}{k_{\rm s}}).
\end{align}
Then $D_tA_t^\mathrm{hk}$ becomes\begin{align}
    D_tA_t^\mathrm{hk}=\mqty(0&-a\\a&0)
\end{align}
and the eigenvalues of $D_tA_t^\mathrm{hk}$ are given by
\begin{align}
    \lambda_1&=a\mathrm{i}, &\quad \lambda_2&=-a\mathrm{i},
\end{align}
which are proportional to the intensity of the rotational force $a$. 

Because the spectral decomposition of $D_tA_t^\mathrm{hk}$ is given by
\begin{align}
    \mathsf{F}_1&=\frac{1}{2}\mqty(1&\mathrm{i}\\-\mathrm{i}&1), &\quad \mathsf{F}_2&=\frac{1}{2}\mqty(1&-\mathrm{i}\\\mathrm{i}&1),
\end{align}
the normalized intensities are given by $J_1 =J_2 = 1/k$ and the decomposition of the housekeeping entropy production rate $\sigma_t^{\mathrm{hk}} = \sum_k \sigma_t^{\mathrm{hk} (k)} =\sum_k |\lambda_k |^2 J_k$ is \begin{align}
\sigma_t^{\mathrm{hk(1)}}=\frac{a^2}{k_{\rm s}}, \hspace{5mm}  \sigma_t^{\mathrm{hk(2)}}=\frac{a^2}{k_{\rm s}}, \hspace{5mm} 
     \sigma_t^{\mathrm{hk}} = \frac{2 a^2}{k_{\rm s}}
\end{align}
We can again confirm that the sum of the contributions of each mode is equal to the housekeeping entropy production rate. This housekeeping entropy production rate is equivalent to the entropy production rate in the steady state, which is proportional to $a^2$.

\subsection{Example 2: Numerical calculation for 2D spring-linked particles with rotational force in the steady state}\label{section_example2}

In the second example, we prepare a system that generates several oscillatory modes with different frequencies. We consider two two-dimensional particles with the positions $(x_1,y_1)^\top$ and $(x_2,y_2)^\top$. The two particles receive rotational forces $(-ay_1,ax_1)^\top$ and $(-by_1,bx_1)^\top$, respectively, and are connected to a spring with a spring constant $k_{\rm s}$, which is fixed at the origin. Additionally, the two particles are connected to each other through a spring with spring constant $k_{\rm s}'$ (Fig. \ref{fig:fig_toy_example}b). The Langevin equation for the system is given by
\begin{align}
    d\mqty(x_1\\y_1\\x_2\\y_2)&=\mqty(
    -k_{\rm s}x_1-k_{\rm s}'(x_1-x_2)-ay_1 \\
    -k_{\rm s}y_1-k_{\rm s}'(y_1-y_2)+ax_1 \\
    -k_{\rm s}x_2-k_{\rm s}'(x_2-x_1)-by_2 \\
    -k_{\rm s}y_2-k_{\rm s}'(y_2-y_1)+bx_2 
    )dt+\sqrt{2}G_td\bm{B}_t,\label{eq:Langevin_example2}
\end{align}
and 
\begin{align}
 G_t=\mqty(\sqrt{T}&0&0&0\\0&\sqrt{T}&0&0\\0&0&\sqrt{T}&0\\0&0&0&\sqrt{T}), \hspace{5mm} D_t=\mqty(T&0&0&0\\0&T&0&0\\0&0&T&0\\0&0&0&T)
\end{align}
where a positive scalar $T$ is the temperature of the heat bath. This equation corresponds to $\bm{x}_t=(x_1, y_1, x_2, y_2)^\top$ and \begin{align}
    D_tA_t&=\mqty(
        -k_{\rm s}-k_{\rm s}'&-a&k_{\rm s}'&0\\
        a&-k_{\rm s}-k_{\rm s}'&0&k_{\rm s}'\\
        k_{\rm s}'&0&-k_{\rm s}-k_{\rm s}'&-b\\
        0&k_{\rm s}'&b&-k_{\rm s}-k_{\rm s}'
    )\\
    D_t\bm{b}_t&=\mqty(0\\0\\0\\0)
\end{align}
in the Langevin equation (\ref{eq:Langevin_linear}). 

In this example, we only consider the steady state and show a numerical calculation. $\bm{\mu}_t$ and $\mathit{\Sigma}_t$ are set to the values that satisfy the steady state. 

For various spring constant $k_{\rm s}'$ values, we have numerically computed the eigenvalues of $D_t A^{\rm hk}_t$ in Fig. \ref{fig:fig_toy_example}c and the corresponding contributions to the housekeeping entropy production rate $\sigma_t^{\mathrm{hk}(k)}$ in Fig. \ref{fig:fig_toy_example}d. These are plotted for the parameters $a=1, b=4, k_{\rm s}=1$. 

When $k_{\rm s}'$ is small, the eigenvalues are close to $\pm a\mathrm{i}$ and $\pm b\mathrm{i}$. The corresponding contributions to the housekeeping entropy production rate are around $a^2/k$ and $b^2/k$, respectively. This behavior is theoretically understood. When $k'_{\rm s}=0$, there is no spring connecting the two particles, which can be considered as a situation wherein the two particles in example 1 are placed independently with different rotational forces. We can then reuse the analytical calculation of example 1. The eigenvalues $\lambda_k$ should then be $\pm a\mathrm{i}$ and $\pm b\mathrm{i}$, and the corresponding contributions to the housekeeping entropy production rate $\sigma_t^{\mathrm{hk}(k)}$ should be $a^2/k_{\rm s}$ for $k=1,2$ and $b^2/k_{\rm s}$ for $k=3,4$, respectively. This statement agrees with the numerical calculation  in Fig. \ref{fig:fig_toy_example}c-d. 

When $k'_{\rm s}$ is sufficiently large, the eigenvalues of the pairs approach each other, and the eigenvalues asymptotically become around $\pm \alpha\mathrm{i}$, where we set $\alpha= (a+b)/2$ in Fig.~\ref{fig:fig_toy_example}c. The corresponding contributions to the housekeeping entropy production rate $\sigma^{{\rm hk}(k)}_t =\alpha^2/(2 k_{\rm s})$ gives the housekeeping entropy production $\sigma^{{\rm hk}}_t=\sum_{k=1}^4 \sigma^{{\rm hk}(k)}_t =2\alpha^2/k_{\rm s}$ in Fig.~\ref{fig:fig_toy_example}d. Again, this behavior is intuitively understood based on the change of variables. Using the center of mass coordinate 
\begin{align}
    m_x=\frac{x_1+x_2}{2}, \hspace{5mm} m_y=\frac{y_1+y_2}{2}
\end{align}
and the relative coordinates
\begin{align}
    r_x=\frac{x_1-x_2}{2}, \hspace{5mm} r_y=\frac{y_1-y_2}{2}, 
\end{align}
the Langevin equation in Eq. (\ref{eq:Langevin_example2}) becomes
\begin{align}
    d\mqty(m_x\\m_y\\r_x\\r_y)=&\mqty(
    -k_{\rm s}&-\alpha&0&-\beta\\
    \alpha&-k_{\rm s}&\beta&0\\
    0&-\beta&-(k_{\rm s}+2k_{\rm s}')&-\alpha\\
    \beta&0&\alpha&-(k_{\rm s}+2k_{\rm s}')
    )\mqty(m_x\\m_y\\r_x\\r_y)dt \nonumber\\
    &+\sqrt{2}G_t'd\bm{B}_t',\label{eq:langevin_centor_relative}
\end{align}
where we use \begin{align}
    \alpha=\frac{a+b}{2}, \hspace{5mm} \beta=\frac{a-b}{2}, \hspace{5mm} G'_t&=G_t\sqrt{2}.
\end{align}
and \begin{align}
    d\bm{B}_t'=\frac{1}{\sqrt{2}}\mqty(dB_1+dB_3\\dB_2+dB_4\\dB_1-dB_3\\dB_2-dB_4).
\end{align}
Note that $d\bm{B}'_t$ becomes a standard 4-dimensional Brownian motion, satisfying $\mathbb{E}[d\bm{B}_t']=0$ and $\mathbb{E}[d\bm{B}_t'(d\bm{B}_s')^\top]=\delta(s-t)Idt$. When $k'_{\rm s}$ is sufficiently large, the relative coordinates $(r_x,r_y)$ are quickly pulled close to the origin $(r_x,r_y) =(0, 0)$ in the steady state, and contributions from $(r_x,r_y) =(0, 0)$ can be ignored. Thus, the Langevin equation for the center of mass coordinate is approximately given by \begin{align}
    d\mqty(m_x\\m_y)=&\mqty(
    -k&\alpha\\
    -\alpha&-k)\mqty(m_x\\m_y)dt+\sqrt{2} G_t' \mqty(dB'_1\\dB'_2),\label{eq:m_approximate}
\end{align}
when $k'_{\rm s}$ is sufficiently large. This equation is regarded as a single particle system, and we can reuse the analytical calculation from Example 1. In eq. (\ref{eq:m_approximate}), since the particle receives a rotational force proportional to $\alpha$, the eigenvalues become $\pm \alpha \mathrm{i}$ and the housekeeping entropy production rate becomes $\sigma^{{\rm hk}}_t =2 \alpha^2/k_{\rm s}$. On the other hand, the degree of the relative coordinates $(r_x,r_y)$ does not contribute to the housekeeping entropy production rate when $k'_{\rm s}$ is sufficiently large because the relative coordinates $(r_x,r_y)$ are always fixed at $(0, 0)$ and the velocity field in the relative coordinates vanishes. We note that the above discussion is based on the invariance of the housekeeping entropy production rate under the linear transformation. For the invariance of the housekeeping entropy production rate under the linear transformation, see Appendix \ref{appendix_invariance_for_linear_transform}.

\section{Application to ECoG Data}

\subsection{Examples of results from a single time window in one recording session of a monkey}\label{section:example_ecog}

To demonstrate the utility of our mode decomposition of the housekeeping entropy production rate, we applied it to neural data and compared the properties of the decomposition during awake and anesthetized conditions. It is well known that the Fourier power of the delta wave (0.5-4 Hz) increases during anesthetized conditions~\cite{Miyasaka1968-zz}. We compared how contributions of each frequency band to the housekeeping entropy production rate vary depending on these conditions with different oscillatory properties. Additionally, we compared the results of our decomposition across different agents used to induce anesthesia, such as ketamine, medetomidine, combination of ketamine and medetomidine, and propofol to study to what degree the effects on the decomposition are common or differ among anesthetic agents. Each anesthetic agent is known to affect the brain differently. For details of the different effects of these anesthetic agents, see Discussion \ref{Discussion_findings} and Ref~\cite{alkire2008consciousness, brown2010general, brown2011general, brown2018multimodal, mashour2024anesthesia}.

We used a 128-channel ECoG open dataset recorded from monkeys~\cite{Nagasaka2011-gk}. Details of the preprocessing and data analysis are explained in the Methods section. After the preprocessing, the number of channels is decreased to 64 since we applied bipolar rereference as preprocessing to remove common artifacts across electrodes. The data were recorded during awake eyes opened, awake eyes closed, and anesthetized conditions. Note that we fit the multivariate time series of ECoG signals to the linear Langevin equation (\ref{eq:Langevin_linear}) with the noise coefficient matrix $D_t$ being a diagonal matrix. We used the linear autoregressive model as a reasonable model based on the previous finding that this model outperforms nonlinear models in terms of r-squared when fitting ECoG time series data \cite{nozari2023macroscopic}.

We divided the data into 60-second time windows and calculated the housekeeping entropy production rate for each. We assume piecewise stationarity for each time window to calculate the housekeeping entropy production rate.  This assumption means that the system is in a steady state only within each 60-second time window, but that the steady state can vary across time windows. This assumption can be justified if the system quickly relaxes to a steady state in response to changes in the environment, and if this steady state slowly changes due to changes in the environment. In our analysis, the housekeeping entropy production rate is equivalent to the entropy production rate because the system is assumed to be in a steady state within each time window. 

\begin{figure}[t]
\centering
\includegraphics[width=\linewidth]{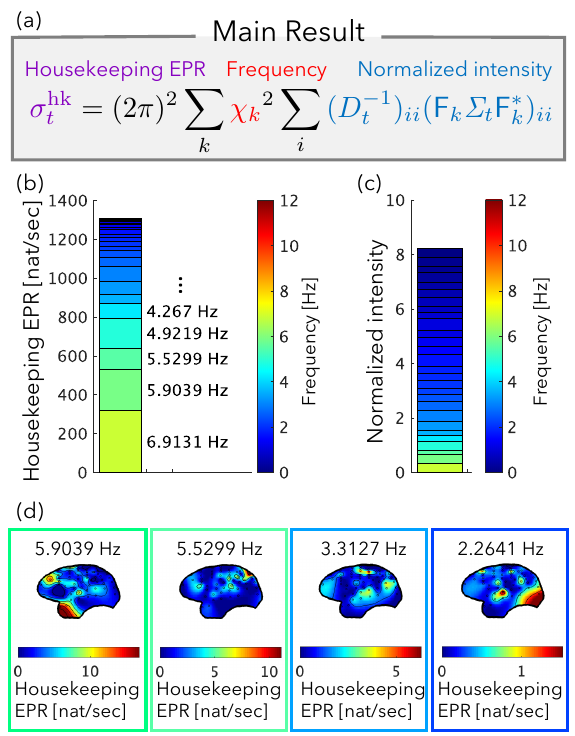}
\caption{
An application example of our decomposition [Eq. (\ref{eq:decomp_oscillation})] to ECoG data. 
(a) Our decomposition of the housekeeping entropy production rate (EPR) [Eq. (\ref{eq:EP_hk_decomposition_uniform})]. The housekeeping entropy production rate can be decomposed into the contributions of each oscillatory mode, which are the products of the squared frequency and the normalized intensity.
(b) An example of our decomposition with respect to each oscillatory  mode. The stacked bar graph shows the contributions of each oscillatory mode $\sigma_t^{\mathrm{hk}(k)}$ in Eq. (\ref{eq:def_sigma_k}). The colors represent the frequency $\chi_k$. The sum of these contributions represents the total housekeeping entropy production rate $\sigma_t^\mathrm{hk}$. 
(c) The normalized intensity $J_k$ [Eq. (\ref{eq:def_J_k})] for comparison. 
(d) Spatial distribution of contributions to the housekeeping entropy production rate for each oscillatory mode $\bm{\sigma}_t^{\mathrm{hk}(k)}$ [Eq. (\ref{eq:def_sigma_vec})]. Each element of $\bm{\sigma}_t^{\mathrm{hk}(k)}$ is represented by the color at the location of each electrode.
Each plot shows the brain viewed from the lateral side. 
}
\label{fig:fig_example_decomp}
\end{figure}
In applications to the neural data analysis in this study, the dimension $i$ corresponds to the electrodes of the ECoG recordings. Since the spatial positions of the electrodes are given, we can consider the decomposition [Eq. (\ref{eq:decomp_oscillation})] as the spatio-temporal decomposition. 
Then, $\sigma_t^{\mathrm{hk}(k,i)}$ is the contribution of the $i$-th electrode of the $k$-th oscillatory mode. 
As will be shown later in Fig. \ref{fig:fig_example_decomp}d, this enable us to plot the spatial distribution of contributions to the housekeeping entropy production rate. 
The spatial distribution of contributions from the $k$-th oscillatory mode is expressed as the vector
\begin{equation}
    \bm{\sigma}_t^{\mathrm{hk}(k)} = (\sigma_t^{\mathrm{hk}(k,i=1)}, \sigma_t^{\mathrm{hk}(k,i=2)}, \cdots, \sigma_t^{\mathrm{hk}(k,i=d)}). \label{eq:def_sigma_vec}
\end{equation}
Note that the decomposition into the elements $\bm{\sigma}_t^{\mathrm{hk}(k)}$ does not always correspond to the spatial patterns, and in ECoG data analysis exploration of the spatial structure is possible since ECoG data has a spatial structure.
We exemplify the application of our decomposition using a certain 60-second time window of the ECoG multivariate time series recorded in a monkey during the awake eyes-closed condition (Fig. \ref{fig:fig_example_decomp}). The result of the application of our decomposition is illustrated in Fig. \ref{fig:fig_example_decomp}b. 
There are 64 modes, including plus and minus frequencies, which is equal to the number of channels. In Fig. \ref{fig:fig_example_decomp}b, the modes with the same absolute values of frequencies are combined and shown as a single mode. Hence, 32 modes are shown in Fig. \ref{fig:fig_example_decomp}b. The exact number of modes is different among individual monkeys depending on the number of electrodes discarded in preprocessing. The stacked bar graph shows contributions from each oscillatory mode $\sigma_t^{\mathrm{hk}(k)}$. The sum of these contributions represents the total housekeeping entropy production rate, $\sigma_t^\mathrm{hk}$.
We observed the amount of contribution to the housekeeping entropy production rate from the 6.9131 Hz component, the 5.9039 Hz component and so on. 

Our decomposition is different from just plotting the normalized intensity of the oscillation $J_k$ in Eq. (\ref{eq:def_J_k}) (Fig. \ref{fig:fig_example_decomp}c). 
Compared to simply plotting the normalized intensity of the oscillation $J_k$ (Fig. \ref{fig:fig_example_decomp}c), $\sigma_t^{\mathrm{hk}(k)}$ of high frequency components are emphasized and $\sigma_t^{\mathrm{hk}(k)}$ of low frequency components are diminished (Fig. \ref{fig:fig_example_decomp}b), because the squares of the frequencies $\chi_k$ are multiplied in $\sigma_t^{\mathrm{hk}(k)}$ [Eq. (\ref{eq:def_sigma_k})]. 

The decomposition in Eq. (\ref{eq:decomp_oscillation}) also allows us to plot the spatial distribution of the contributions to the entropy production rate from each oscillatory mode $\bm{\sigma}_t^{\mathrm{hk}(k)}$ (Fig. \ref{fig:fig_example_decomp}d) [Eq. (\ref{eq:def_sigma_vec})]. This is because the decomposition is not only a decomposition into oscillatory elements but also a decomposition into the contributions from each dimension of the spatial oscillatory mode [Eq. (\ref{eq:decomp_oscillation})].  In Fig. \ref{fig:fig_example_decomp}d, each element of $\bm{\sigma}_t^{\mathrm{hk}(k)}$ was represented by the color at the location of each electrode, and the values between electrode positions were interpolated using the \texttt{ft\_topoplotER} function from FieldTrip~\cite{oostenveld2011fieldtrip}.

\subsection{Temporal stability of the decomposition}\label{section:temporal_stability_ecog}
\begin{figure}
\centering
\includegraphics[width=\linewidth]{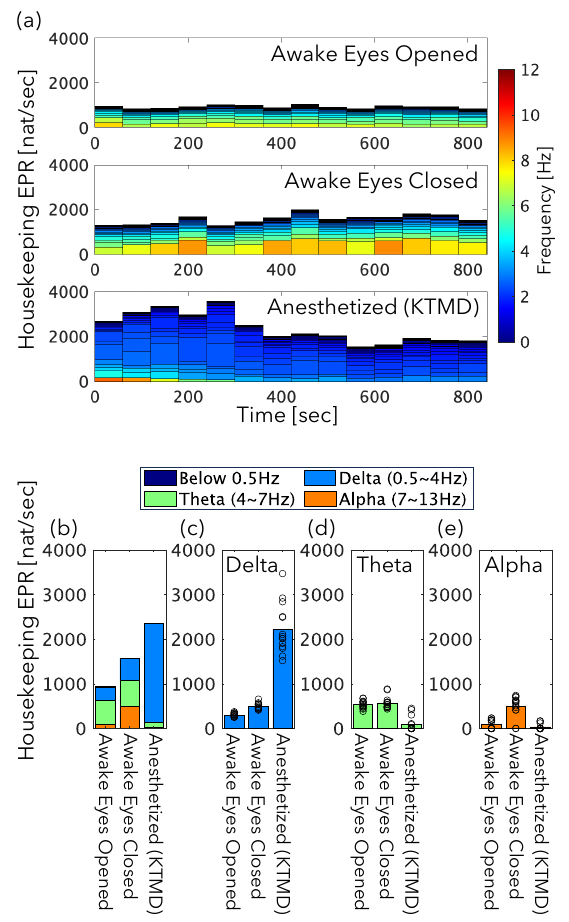}
\caption{
The contributions of each frequency band to the housekeeping entropy production rate (EPR) were stable across time windows compared to the differences across awake and ketamine-medetomidine-induced anesthetized conditions. 
(a) $\sigma_t^{\mathrm{hk}(k)}$ from each 60 second time window are shown. The color indicates the frequency $\chi_k$. 
(b-e) The contribution to the housekeeping entropy production rate from the delta band $\sigma_t^{\mathrm{hk, Delta}}$, the theta band $\sigma_t^{\mathrm{hk, Theta}}$ and the alpha band $\sigma_t^{\mathrm{hk, Alpha}}$ [Eqs. (\ref{eq:def_sigma_delta}), (\ref{eq:def_sigma_theta}), and (\ref{eq:def_sigma_alpha})] are shown. 
Stacked bar plots of all frequency bands are shown in (b). The displayed values are the averages over time windows. In (c-e), the contributions to the housekeeping entropy production rate from (c) the delta band $\sigma_t^{\mathrm{hk, Delta}}$, (d) the theta band $\sigma_t^{\mathrm{hk, Theta}}$, and (e) the alpha band $\sigma_t^{\mathrm{hk, Alpha}}$ are respectively shown. The bar plots represent the averaged values over time windows. Each circle represents the value of each time window. 
}
\label{fig:time_variatiom}
\end{figure}

Before investigating the difference across conditions, we first assessed how stable the decomposition of the housekeeping entropy production rate  $\sigma_t^{\mathrm{hk}(k)}$ is over time
(Fig. \ref{fig:time_variatiom}).
Here, as demonstration, we only show the results from the ketamine-medetomidine anesthetic condition. 
We found that the proportion of contribution to the housekeeping entropy production rate from each frequency $\sigma_t^{\mathrm{hk}(k)}$ was stable over time under all conditions (Fig. \ref{fig:time_variatiom}a). 
We plotted $\sigma_t^{\mathrm{hk}(k)}$ of each 60-second time window. We observed that although there were small time variations in $\sigma_t^{\mathrm{hk}(k)}$, the time variations were much smaller than the difference across conditions.

To assess whether the degree of temporal variation in each frequency's contribution to the housekeeping entropy production rate is small enough to detect the difference in each frequency's contribution across conditions, we binned the oscillatory modes into the delta band (0.5-4Hz), theta band (4-7Hz), alpha band (7-13Hz) and the frequency band below 0.5Hz according to the conventional definition of frequency bands in the brain \cite{kandel2000principles} (Fig. \ref{fig:time_variatiom}b).
The sums of the contributions from each frequency band are summarized as $\sigma_t^\mathrm{hk, <0.5}$, $ \sigma_t^\mathrm{hk, Delta}$, $\sigma_t^\mathrm{hk, Theta}$ and $ \sigma_t^\mathrm{hk, Alpha}$ [Eqs. (\ref{eq:def_sigma_05}), (\ref{eq:def_sigma_delta}), (\ref{eq:def_sigma_theta}), and (\ref{eq:def_sigma_alpha}); see Methods section]. The sums of the contribution from each frequency band $ \sigma_t^\mathrm{hk, <0.5}$, $ \sigma_t^\mathrm{hk, Delta}$, $ \sigma_t^\mathrm{hk, Theta}$, and $ \sigma_t^\mathrm{hk, Alpha}$ calculated from each time window are shown in dot plots, and their averages across time windows are shown in bar plots (Fig. \ref{fig:time_variatiom}c-e). 
The plots show that the contributions to the housekeeping entropy production rate from each frequency band $\sigma_t^\mathrm{hk, Delta}$, $\sigma_t^\mathrm{hk, Theta}$, and $\sigma_t^\mathrm{hk, Alpha}$ differed across conditions. 
In this single monkey, we observed that the contributions of the delta band $\sigma_t^\mathrm{hk, Delta}$ were larger in the anesthetized condition than in the awake conditions; contributions of the theta band $\sigma_t^\mathrm{hk, Theta}$ were larger in the awake conditions than in the anesthetized condition; and the contributions of the alpha band $\sigma_t^\mathrm{hk, Alpha}$ were larger in the awake eyes-closed condition than in the awake eyes-open condition or in the anesthetized condition. From the dot plots in Figs. \ref{fig:time_variatiom}c-e, we can see that the time variations of $\sigma_t^\mathrm{hk, Delta}$, $ \sigma_t^\mathrm{hk, Theta}$, and $\sigma_t^\mathrm{hk, Alpha}$ were smaller than the differences of $\sigma_t^\mathrm{hk, Delta}$, $ \sigma_t^\mathrm{hk, Theta}$, and $\sigma_t^\mathrm{hk, Alpha}$ across the conditions. For the other monkeys, we also found that there are similar trends in the differences of $\sigma_t^\mathrm{hk, Delta}$ and $ \sigma_t^\mathrm{hk, Theta}$ across the conditions, and that the temporal variations are smaller than these differences across the conditions, i.e., the decomposition is temporally stable (Fig. S1).

\begin{figure*}
\centering
\includegraphics[width=0.8\linewidth]{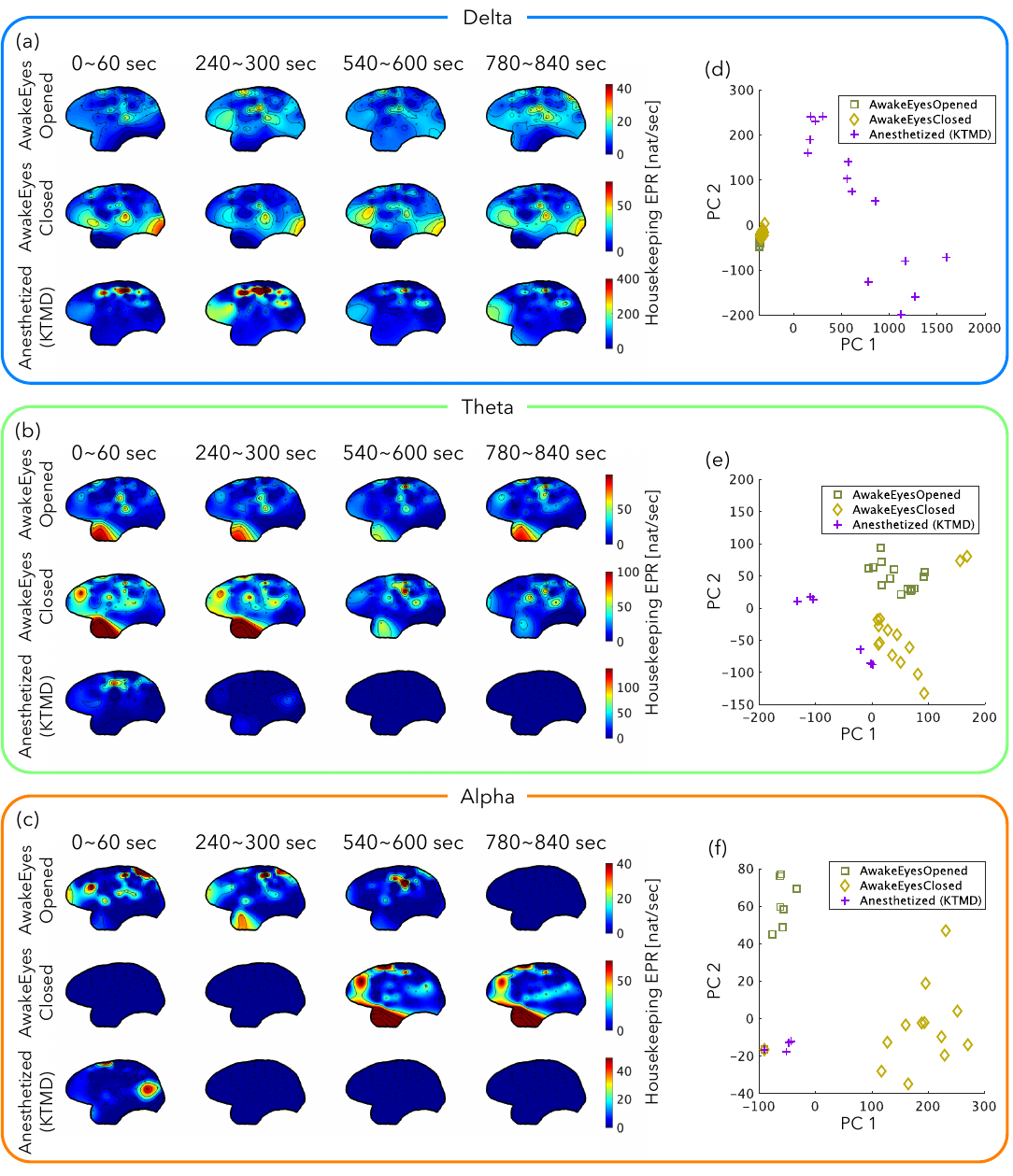}
\caption{
The spatial distributions of contributions from each frequency band to the housekeeping entropy production rate were stable across time windows compared to the differences across  awake and ketamine-medetomidine-induced anesthetized conditions. 
(a-c) The spatial distributions of contributions to the housekeeping entropy production rate from (a) the delta band $\bm{\sigma}_t^{\mathrm{hk, Delta}}$, (b) the theta band $\bm{\sigma}_t^{\mathrm{hk, Theta}}$, and (c) the alpha band $\bm{\sigma}_t^{\mathrm{hk, Alpha}}$ from four example time windows [Eqs. (\ref{eq:def_sigma_delta_vec}), (\ref{eq:def_sigma_theta_vec}), and (\ref{eq:def_sigma_alpha_vec})]. 
Note that for frequency bands that were not detected as an eigenvalue $\lambda_k$ of the matrix $\mathit{D}_t A_t^\mathrm{hk}$, zero values were plotted (the blue color was uniformly plotted) because the contributions of such components were 0. 
The quantities $\bm{\sigma}_t^{\mathrm{hk, Delta}}$, $\bm{\sigma}_t^{\mathrm{hk, Theta}}$, and $\bm{\sigma}_t^{\mathrm{hk, Alpha}}$ are represented by the colors at the location of each electrode.
Each plot shows the brain viewed from the lateral side. 
(d-f) The spatial distributions of the contributions to the housekeeping entropy production rate from (d) the delta band $\bm{\sigma}_t^{\mathrm{hk, Delta}}$, (e) the theta band $\bm{\sigma}_t^{\mathrm{hk, Theta}}$, and (f) the alpha band $\bm{\sigma}_t^{\mathrm{hk, Alpha}}$ were visualized by projecting them onto two dimensions using principal component analysis. 
In the scatter plots, each point corresponds to the spatial distribution from each time window, projected onto two principal component spaces.
}
\label{fig:topoplot_time_variation}
\end{figure*}

Furthermore, we found that the spatial distributions of the contributions to the entropy production rate from each frequency band were also stable across time windows (Fig. \ref{fig:topoplot_time_variation}). 
The spatial distributions of the contributions to the entropy production rate from each frequency band are defined as the sums of $\bm{\sigma}_t^{\mathrm{hk}(k)}$ within each frequency band, namely as $\bm{\sigma}_t^{\mathrm{hk, Delta}}$, $\bm{\sigma}_t^{\mathrm{hk, Theta}}$, and $\bm{\sigma}_t^{\mathrm{hk, Alpha}}$ [Eqs. (\ref{eq:def_sigma_delta_vec}), (\ref{eq:def_sigma_theta_vec}), and (\ref{eq:def_sigma_alpha_vec}); see Methods section]. Figure  \ref{fig:topoplot_time_variation}a-c shows the spatial distributions $\bm{\sigma}_t^{\mathrm{hk, Delta}}$, $\bm{\sigma}_t^{\mathrm{hk, Theta}}$, and $\bm{\sigma}_t^{\mathrm{hk, Alpha}}$ from four example time windows. 
We observed that the spatial distributions $\bm{\sigma}_t^{\mathrm{hk, Delta}}$, $\bm{\sigma}_t^{\mathrm{hk, Theta}}$, and $\bm{\sigma}_t^{\mathrm{hk, Alpha}}$ were stable over time under all conditions. 

To assess whether the degree of temporal variation in the spatial distribution is small enough to detect the difference in spatial distribution across conditions, the vectors of the spatial distributions of the contributions from each frequency band $\bm{\sigma}_t^{\mathrm{hk, Delta}}$, $\bm{\sigma}_t^{\mathrm{hk, Theta}}$, and $\bm{\sigma}_t^{\mathrm{hk, Alpha}}$, whose dimension is the number of electrodes, were visualized by projecting them onto two dimensions using principal component analysis (Fig. \ref{fig:topoplot_time_variation}d-f). 
In the scatter plots, each point corresponds to the spatial distribution from each time window, projected onto two principal component spaces. 
In the delta band, points corresponding to the awake eyes-opened and awake eyes-closed conditions are plotted at closer positions, while those corresponding to the anesthetized condition are plotted at more distant positions. For the theta and alpha bands, points are plotted in separate positions for each condition. In all instances, the plots show that the variations between time windows are smaller than the differences across conditions.
Note that for frequency bands not detected as an eigenvalue $\lambda_k$ of the matrix $\mathit{D}_t A_t^\mathrm{hk}$, zero values are plotted in Fig. \ref{fig:topoplot_time_variation}a-c, as observed for example in the plot for the alpha waves under anesthetic conditions. Such oscillatory modes are projected to the same point, as can be seen at the bottom right of Fig. \ref{fig:topoplot_time_variation}f.

\subsection{Comparison of the results of our decomposition across awake and different anesthetized conditions}\label{section:comparison_ecog}
\begin{figure*}[t]
\centering
\includegraphics[width=\linewidth]{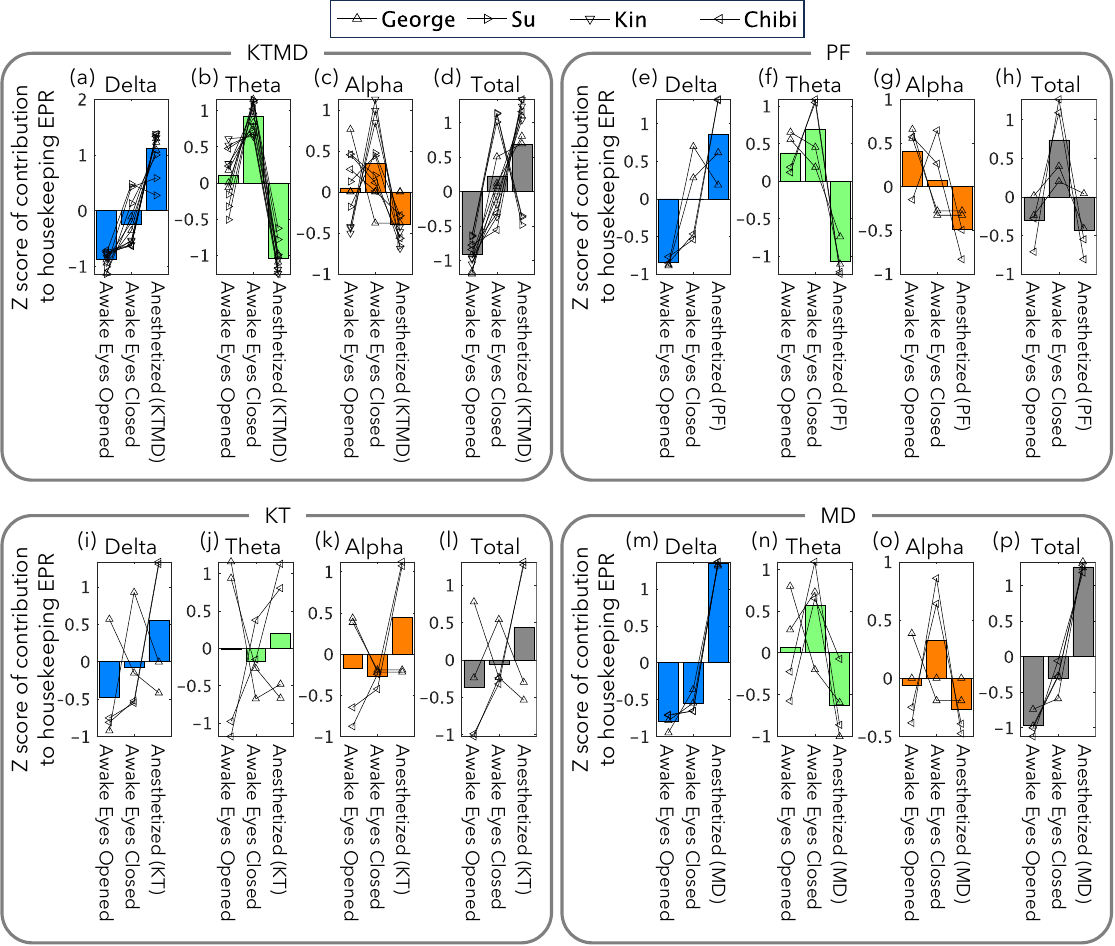}
\caption{
Difference in the contribution of each frequency band to the housekeeping entropy production rate (EPR) between different conditions for all monkeys and recording sessions.
(a-c) The plots show the z-scores of the contribution to the housekeeping EPR from (a) the alpha band $\sigma_t^{\mathrm{hk, Delta}}$, (b) theta band $\sigma_t^{\mathrm{hk, Theta}}$, and (c) alpha band $\sigma_t^{\mathrm{hk, Alpha}}$ during awake and ketamine-medetomidine-induced anesthetized conditions. Each dot represents z-scores calculated in a single recording session from a single monkey. The displayed values are the averages of the z-scores across time windows within each condition. In the plot, results from the same recording session from the same monkey are connected by lines, and markers represent individual monkeys (George, Su, Kin and Chibi). Bar plots show the average across monkeys and recording sessions. 
(d) The z-scores of the housekeeping EPR $\sigma_t^\mathrm{hk}$ are shown in a similar way to (a-c). 
(e-p) Same as (a-d), but (e-h) propofol, (i-l) ketamine, and (m-p) medetomidine were used for the anesthetized conditions. 
}
\label{fig:group_KTMD}
\end{figure*}

Having confirmed that the contributions from each frequency band $\sigma_t^{\mathrm{hk, Delta}}$, $\sigma_t^{\mathrm{hk, Theta}}$, and $\sigma_t^{\mathrm{hk, Alpha}}$ were stable over time (Figs. \ref{fig:time_variatiom} and \ref{fig:topoplot_time_variation}), we investigated whether there were consistent features of the decomposition for different conditions across multiple monkeys with multiple recording sessions. 
Figure \ref{fig:group_KTMD} summarizes the results of decomposition across individual monkeys and recording sessions. For each monkey and session, we computed the z-scores of $\sigma_t^{\mathrm{hk, Delta}}$, $\sigma_t^{\mathrm{hk, Theta}}$, and $\sigma_t^{\mathrm{hk, Alpha}}$ and compared them across conditions (Fig. \ref{fig:group_KTMD}a-c).
We calculated their z-scores across all time windows and conditions independently for each monkey, recording session, and frequency band. 
For details of the calculation of z-scores, see the Methods section. 
We also calculated the z-scores of $\sigma_t^\mathrm{hk}$ across all time windows and conditions independently for each monkey in each recording session (Fig. \ref{fig:group_KTMD}d).

\begin{figure*}[t]
\centering
\includegraphics[width=\linewidth]{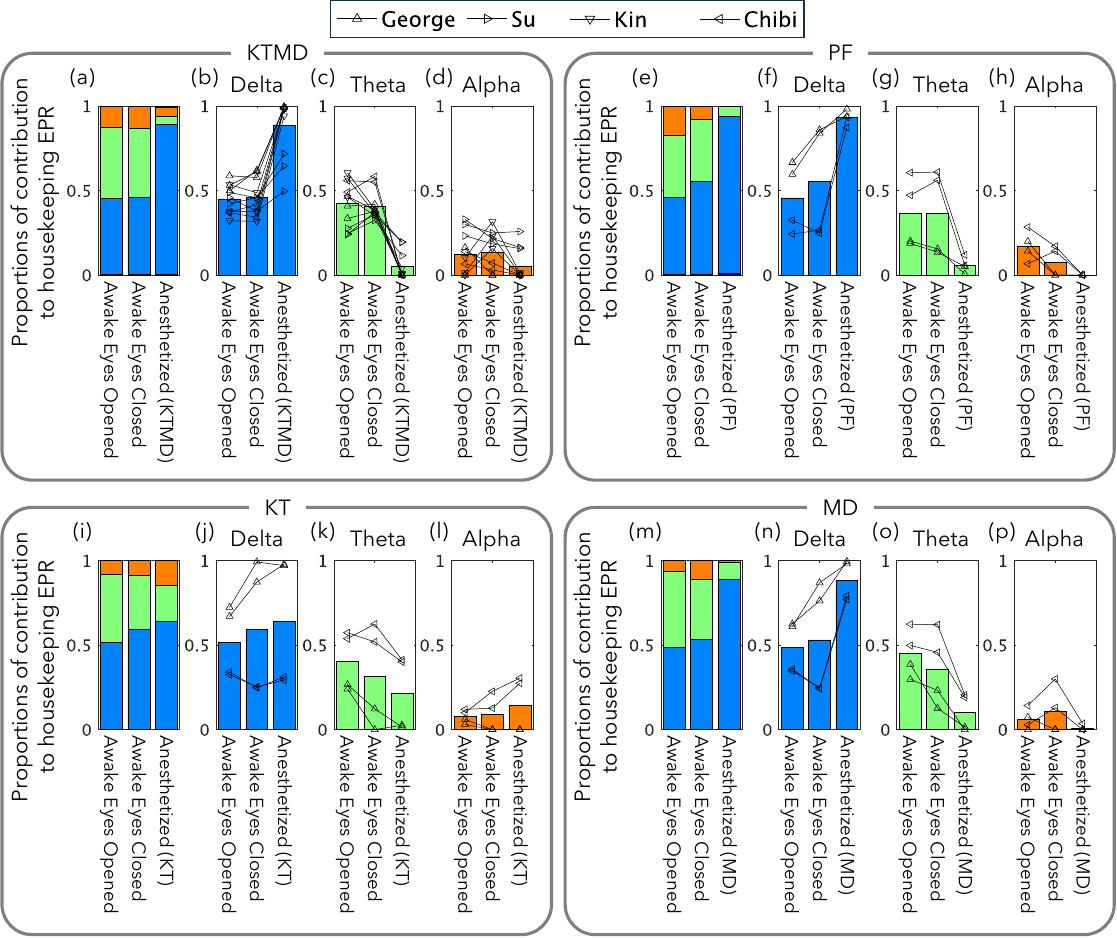}
\caption{Difference in the proportion of contribution of each frequency band to the housekeeping entropy production rate (EPR) between different conditions for all monkeys and recording sessions. (a) The plots show the proportions of the contribution to the housekeeping EPR from the alpha band $\sigma_t^{\mathrm{hk, Delta}}$, the theta band $\sigma_t^{\mathrm{hk, Theta}}$, and the alpha band $\sigma_t^{\mathrm{hk, Alpha}}$ during awake and ketamine-medetomidine-induced anesthetized conditions. The averages across monkeys and recording sessions are shown. (b-d) The results for (b) the alpha band $\sigma_t^{\mathrm{hk, Delta}}$, (c) the theta band $\sigma_t^{\mathrm{hk, Theta}}$, and (d) the alpha band $\sigma_t^{\mathrm{hk, Alpha}}$ are separately shown. Each dot represents the proportion calculated in a single recording session from a single monkey. The displayed values are the averages of the proportion across time windows within each condition. In the plot, results from the same recording session from the same monkey are connected by lines, and markers represent individual monkeys (George, Su, Kin and Chibi). Bar plots show the average across monkeys and recording sessions. (e-p) Same as (a-d), but (e-h) propofol, (i-l) ketamine, and (m-p) medetomidine are used for the anesthetized conditions.}
\label{fig:group_KTMD_prop}
\end{figure*}

Before explaining all of the results obtained across different anesthetics, we first explain in detail an example result from the ketamine-medetomidine-induced anesthetized condition to provide guidance on interpreting the figures. Comparing the awake and ketamine-medeomidine-induced anesthetized conditions, we observed several robust tendencies across all individual monkeys, namely that (i) the contributions of the delta band were larger in the anesthetized condition than in the awake conditions (Fig. \ref{fig:group_KTMD}a), (ii) the contributions of the theta band were smaller in the anesthetized condition than in the awake conditions (Fig. \ref{fig:group_KTMD}b), (iii) the contributions of the alpha band were larger in the awake eyes-closed condition than in the anesthetized condition (Fig. \ref{fig:group_KTMD}c). 
In addition, we observed a modest tendency in 3 out of 4 individual monkeys (iv) that the z-values of the total $\sigma_t^\mathrm{hk}$ were larger in the anesthetized condition than in the awake conditions (Fig. \ref{fig:group_KTMD}d).

Next, we compared the above results across different anesthetic agents to study to what extent the effects on the decomposition are common or differ among anesthetic agents. The results of propofol-, ketamine-, and medetomidine-induced anesthetized conditions are shown in Fig.\ref{fig:group_KTMD}e-p. (i) The larger contribution of the delta band in the anesthetized conditions than in the awake condition, (ii) the smaller contribution of the theta band in the anesthetized conditions than in the awake condition, and (iii) the smaller contribution of the alpha band in the anesthetized conditions than in the awake eyes-closed condition and were robust across all anesthetic agents, except for the ketamine-induced anesthetized condition. (iv) Changes in total housekeeping entropy production rate were not consistent across anesthetic agents. In the ketamine-medetomidine- and medetomidine-induced anesthetized conditions, the housekeeping entropy production rate were larger than in the awake conditions. However, in the propofol-induced anesthetized conditions, the housekeeping entropy production rate was smaller than in the awake eyes-closed conditions. In the ketamine-induced anesthetized condition, there were no consistent trends.  It is important to note that the data from the ketamine-only condition were recorded from only two unique monkeys, while the data from the other conditions were recorded from four unique monkeys. Consequently, the results from the ketamine-only condition are less reliable than those of the other conditions due to the smaller sample size.

To clearly demonstrate the difference in contribution rates to the entropy production rate from different oscillatory modes between the awake and anesthetized conditions, we also showed the proportions of the contributions of each frequency band to the total housekeeping entropy production rate in both the awake and anesthetized conditions in Fig. \ref{fig:group_KTMD_prop}. We found that in awake conditions, delta waves contribute about half of the total housekeeping entropy production rate, while theta and alpha waves account for the other half. In contrast, in the anesthetized condition, delta waves dominated the housekeeping entropy production rate for most anesthetics except ketamine. For the ketamine-induced anesthetized condition, there was no consistent trend across individuals.

\section{Discussion}
In this paper, we derived a relationship between oscillations and the entropy production rate. The housekeeping entropy production rate can be simultaneously decomposed into contributions from each oscillatory mode and each dimension $\sigma_t^{\mathrm{hk}(k,i)}$ [Eq. (\ref{eq:decomp_oscillation})]. Oscillatory modes with a larger normalized intensity $J_k$ or frequency $\chi_k$ make a larger contribution to the housekeeping entropy production rate. 
Furthermore, when applied to neural activity data recorded using ECoG \cite{Nagasaka2011-gk}, it is observed that contributions from each frequency band remain stable. Under anesthetized conditions, compared to the awake conditions, the contributions from the delta wave (0.5-4Hz) were larger, while those from the theta wave (4-7Hz) were smaller.
Since the entropy production rate determines various limits of information processing \cite{Parrondo2015-wo,barato2015thermodynamic,lan2012energy,Nakazato2021-do}, these results might lead to a better understanding of the role of oscillations in brain information processing.

\subsection{Significance of decomposing EPR to oscillatory contributions}\label{discussion:significance}
First, we discuss relationships between our decomposition and other methods for extracting oscillatory components, such as the Fourier transform and dynamic mode decomposition (DMD) \cite{schmid2010dynamic,kutz2016dynamic}. The Fourier transform is a commonly used method in neuroscience that extracts all frequency components below the Nyquist frequency for each electrode. However, the Fourier transform has limitations: (i) it extracts all frequencies indiscriminately, potentially leading to statistical overfitting, and (ii) it does not capture the oscillatory patterns across multiple electrodes. DMD addresses these issues by decomposing the multivariate time series of $\bm{x}$ into a small number of spatio-temporal oscillatory modes through the eigenvalue decomposition of auto-regressive coefficients. The oscillatory modes obtained in our decomposition in Eq. (\ref{eq:oscillation}) are similar to the oscillatory modes obtained in DMD. The important and distinctive feature is that our decomposition provides the thermodynamic meaning for such DMD-like oscillatory modes by elucidating the contributions of each oscillatory mode to the entropy production rate, which we believe is useful for understanding the nature of information processing in the brain.

Specifically, the entropy production rate has utility in determining various physical limits of information processing, including limits of the speed~\cite{aurell2012refined,van2021geometrical, Nakazato2021-do,yoshimura2023housekeeping} and accuracy~\cite{barato2015thermodynamic, seifert2019stochastic, horowitz2020thermodynamic} of the information processing. For potential applications in neuroscience, these relations allow us to quantify, for example, (i) the limits of information processing in the brain, and (ii) how close the brain's information processing is to the physical limits. Such quantifications could provide a mathematical framework to capture, for example, brain development and learning. In such applications, our decomposition may explain previous reports of changes in neural oscillations, such as delta and beta waves, associated with development~\cite{uhlhaas2009development, thatcher2008development} or learning~\cite{pollok2014changes,bauer2007gamma}. As another example, a previous study computed the entropy production rate and discussed the relationships to cognitive cost ~\cite{lynn2021broken}. Our decomposition may be useful for further interpreting the relationships between cognitive cost and the entropy production rate in terms of oscillations. 

We also note its applicability to other systems. Regarding the fluctuation-response relation violation, the heat for Langevin systems can be decomposed via the Fourier transform~\cite{harada2005equality}. In the steady state, the entropy production rate is given by the heat, and the result in Ref.~\cite{harada2005equality} can be generalized for the entropy production rate. Specifically, such a decomposition of the entropy production rate has been discussed in efforts to analyze the spatiotemporal dissipation for active matter~\cite{nardini2017entropy} and nonequilibrium crystals~\cite{caprini2023entropy}. Thus, our DMD-like decomposition may represent a complementary method to analyze spatiotemporal dissipation in other systems, such as the active matter and nonequilibrium crystals. A relation between our DMD-like decomposition and a geometric decomposition via the spatial Fourier transform~\cite{nagayama2023geometric} is also interesting. 

\subsection{Interpretation of the entropy production rate calculated from ECoG recording}

In this section we explain the interpretation of the entropy production rate calculated from the ECoG data. Although it is possible to define the entropy production rate for general Langevin systems including the neural ECoG recording, it is not always directly related to physical heat. General Langevin systems for the macroscopic coarse-grained signal allow us to define a potential force that provides effective dynamics, but there is no guarantee that this potential can actually be regarded as physical energy. The noise intensity for general Langevin systems is also not directly related to the thermal noise in the environment.
Thus, the entropy production rate is a statistical measure of the irreversibility of dynamics for a macroscopic coarse-grained signal, which may not be directly related to physical heat at the microscopic scale. 

In general, the entropy production rate for macroscopic systems can be a lower bound on the entropy production rate at more microscopic scales \cite{esposito2012stochastic}. Therefore, 
the entropy production rate for a macroscopic coarse-grained signal may be an underestimation of the entropy production rate in microscopic scale, which is related to physical heat.  However, ECoG data are not simply coarse-grained representations of microscopic signals, a fact attributable to several factors including the nature of the recording (not a direct recording of neural activity), recording noise, and pre-processing. Consequently, the entropy production rate derived from ECoG data may not provide a direct lower bound for the entropy production rate at microscopic scale. 

Nevertheless, the entropy production rate has a physical meaning even for macroscopic coarse-grained signals. Various thermodynamic trade-off relations for general Langevin systems~\cite{aurell2012refined, gingrich2017inferring, dechant2018current, ito2020stochastic, otsubo2020estimating, koyuk2020thermodynamic, Nakazato2021-do, dechant2022geometric,ito2023geometric} provide physical meaning for the entropy production rate even for macroscopic coarse-grained signals. Based on thermodynamic trade-off relations, the entropy production rate for macroscopic systems provides the physical limits for speed and fluctuation observable at the corresponding scale. The thermodynamic trade-off relations still hold for time series of ECoG data, and thus the entropy production rate may be significant as a physical limit for information processing at the macroscopic scale. Indeed, neuroscience research has identified various insights at this macroscopic level, extracting information related to visual~\cite{majima2014decoding} or auditory experiences~\cite{ramsey2018decoding}, working memory~\cite{van2013decoding}, motor movements~\cite{pistohl2012decoding}, and intentions~\cite{spuler2014decoding} based on statistics of ECoG data. Since the entropy production rate for the macroscopic ECoG recordings, as a statistical measure of irreversibility, provides a physical limit for information processing at the same macroscopic level, we believe that the application of stochastic thermodynamics to these domains opens avenues for investigating how such diverse insights in neuroscience are explained in terms of thermodynamic limits. Our theoretical results [Eq. (\ref{eq:decomp_oscillation})] also allow the exploration of thermodynamic effects in the context of neural oscillation.

\subsection{Differences in decomposed entropy production rates between awake and anesthetized conditions}\label{Discussion_findings}

In this section, we discuss the biological or clinical implications of how different oscillatory modes contribute to the entropy production rate in the awake and anesthetized conditions. First, we found that in the anesthetized state, the entropy production rate originates mainly from the delta-band wave, and its contribution is very large, comparable to the total entropy production rate in the awake state. This might be considered a predictable consequence, given previous findings that delta wave power increases and higher frequency oscillations decrease under anesthesia. However, this is not obvious since our theory shows that low frequency oscillations contribute less to the entropy production rate than high frequency oscillations by a factor inversely proportional to the square of their frequency ratio.  Therefore, the increase in delta power must be sufficiently large to overcome this square of the frequency ratio. For example, for a 2 Hz oscillation to contribute more to the entropy production rate than a 6 Hz oscillation, the intensity of the 2 Hz oscillation would need to be more than 9 times larger. Our analysis of the empirical data with the novel decomposition method allows us to quantitatively confirm that the increase in delta power is indeed the primary driver of the high entropy production rate observed during anesthesia. This somewhat counterintuitive result—that anesthesia dominated by a low-frequency wave, theoretically expected to cause a low entropy production rate, causes a large entropy production rate in the empirical data—sheds new light on the biological and clinical consequences of anesthesia.

Although this finding is robust across the different anesthetic agents we investigated, we also found that the ketamine-only condition is subtly different from other anesthetics in terms of contribution rates. In the ketamine-only condition, although the contribution of the delta wave is commonly highest across monkeys, the contributions of the theta and alpha waves  differ. In one monkey, theta and alpha wave contribution was still relatively high, with contribution rates close to those in the awake condition. This may possibly reflect the different effects of each anesthetic agent on neural activity. For example, medetomidine and propofol act primarily as agonists of alpha 2-adrenoceptor and GABA-A receptors, respectively~\cite{brown2010general, brown2011general, brown2018multimodal, alkire2008consciousness, mashour2024anesthesia}, which suppress neural activity. In contrast, ketamine acts as an antagonist of NMDA receptors~\cite{brown2010general, brown2011general, brown2018multimodal, alkire2008consciousness, mashour2024anesthesia}, leading to a dissociative state that induces unresponsiveness, but not necessarily unconsciousness, as evidenced by reports of hallucinations and dream-like experiences during ketamine-induced general anesthesia~\cite{bonhomme2016resting, li2016ketamine}.
For details of the biological effects of each anesthetic agent, see Refs. \cite{brown2010general, brown2011general, brown2018multimodal, alkire2008consciousness, mashour2024anesthesia}. The uniqueness of the ketamine-only condition might also be related to existing findings \cite{deco2022insideout, sarasso2015consciousness, li2016ketamine} which also show the unique behavior of the ketamine-only condition with different measures compared to other anesthetic agents. However, also note that our results for the ketamine-only condition are only from two unique monkeys and are accordingly less reliable than those for the other conditions. Thus, to reliably show the difference between the ketamine-only and the other conditions, additional data are needed.

On the other hand, during the awake state, the entropy production rate is generated by the faster oscillations, such as theta and alpha waves, while the contribution of the delta wave is much lower than in the anesthetized state. Although we did not observe higher frequency oscillations, such as beta waves (13-30Hz) or gamma waves (30-70Hz), we speculate that the entropy production rates in these frequencies are generated as needed, depending on the demands of the particular task, because it is known that these frequencies are enhanced only during specific cognitive tasks such as motor control or visual processing. 
 
Taken together, by using our decomposition, we found for the first time that both the awake and anesthetized states generate a comparable amount of entropy production rate, but that the origin of the entropy production rate -- namely how the entropy production rate is generated from the characteristics of dynamical systems -- drastically differs between the anesthetized and awake states. In the anesthetized state, the entropy production rate is caused by slow- and high-intensity oscillations, whereas in the awake state it is caused by faster- and lower-intensity oscillations.

This difference in the origins of the entropy production rate between the awake and anesthetized states would also imply that the neural mechanisms generating the entropy production rate are different. In fact, it is known that the delta, theta, and alpha waves are generated by different neural mechanisms: the delta wave arises from the interaction of neurons within the thalamocortical circuit~\cite{Miyasaka1968-zz, Buzsaki2006-se}; the theta wave arises from population activity within the hippocampus~\cite{Buzsaki2006-se, buzsaki2002theta}; and the alpha wave arise from the supragranular layers of the cortex and propagate from higher areas to lower areas and the thalamus~\cite{halgren2019generation}. Although it is impossible to investigate these circuits involving deep brain regions using ECoG data, which is derived from the surface of the cortex, applying our decomposition to data in these circuits using deep recordings such as neuropixels will advance our understanding of thermodynamic dissipation and the neural substrates that cause it. Such an understanding cannot be achieved by simply computing the total entropy production rate without our oscillatory mode decomposition.

In light of these findings, our study opens new avenues for clinical applications, particularly in the diagnosis and understanding of disorders of consciousness. Prior research, including studies using monkey ECoG~\cite{Sanz_Perl2021-sc} and human fMRI~\cite{gilson2023entropy}, has highlighted differences in entropy production rates across various states of consciousness, such as awake, anesthetized, and sleep states. Moreover, recent work has shown distinct patterns in the irreversibility of brain dynamics, which is closely tied to entropy production rate, between the awake state and various disordered states of consciousness, including minimally conscious states and unresponsive wakefulness syndrome~\cite{g2023lack}. Importantly, these disorders have also been associated with specific neural oscillation patterns~\cite{piarulli2016eeg, wislowska2017night, koch2016neural}. By integrating our findings on the entropy production rate, particularly those on the differences in origin and characteristics between awake and anesthetized states, with these previous studies, our decomposition approach offers a novel and comprehensive framework for the assessment of consciousness disorders. This framework, by simultaneously considering entropy production rates, oscillation patterns, and their spatial distribution, holds the potential to provide more nuanced and informative diagnostics than with existing methods. Such advances could significantly enhance our understanding and ability to effectively diagnose and treat disorders of consciousness.

\subsection{Differences in total entropy production rates between awake and anesthetized conditions}

Although our main findings are the decomposition of entropy production rates, as discussed in the previous section, here we also discuss the difference in the total entropy rates between awake and anesthetized conditions. We found that the entropy production rate was higher under ketamine-medetomidine-induced or medetomidine-induced anesthetized conditions than under awake conditions in most monkeys (Fig. \ref{fig:group_KTMD}d, p). This result may not be conclusive because our results differ from previous studies that also estimated the entropy production rate \cite{Sanz_Perl2021-sc} or a related measure \cite{deco2022insideout} in the same ECoG data, which are reviewed in Ref. \cite{kringelbach2024thermodynamics}.

For example, using a measure of irreversibility based on the degree of asymmetry in cross-covariances, Ref~\cite{deco2022insideout} reported lower irreversibility in the anesthetized condition for most anesthetic agents and higher irreversibility in the ketamine-induced anesthetized condition compared to awake conditions, which represent a different tendency to that seen in our results (Fig. 6). Although the measure computed in \cite{deco2022insideout} could be related to the entropy production rate to some extent, in the sense that both measures quantify the degree of irreversibility, these measures are nevertheless mathematically distinct. Thus, the discrepancy between our study and the previous study is likely due to the difference in the measures used. A major difference is that the measure of irreversibility in \cite{deco2022insideout} is essentially a pairwise measure calculated from a pair of electrodes, and the degree of irreversibility from all electrodes is quantified as the sum of the pairwise measures. In contrast, the entropy production rate in our study is computed based on the joint distribution of all the electrodes. Therefore, we consider that direct comparison between our study and the previous study is difficult. 

On the other hand, by estimating the entropy production rate, Ref~\cite{Sanz_Perl2021-sc} reported a lower entropy production rate in the anesthetized condition regardless of anesthetic agent than in the awake eyes-closed condition, which represents the opposite tendency of our results. This discrepancy might be due to different statistical modeling used. We treated the brain state as a real-valued vector $\bm{x}_t$ (a continuous state) and computed the entropy production rate by approximating the neural dynamics to a linear Gaussian process, while the previous study binned the low-dimensional brain state extracted by PCA into discrete states and computed the entropy production rate assuming a nonlinear process for the time series of these discrete states. While this approach can account for the nonlinearity of neural dynamics, it cannot capture high-dimensional dynamics and loses the spatial information of the ECoG electrodes.

Further investigation is necessary to potentially resolve discrepancies with previous studies and provide more conclusive insights into the differences in total entropy production rates between awake and anesthetized conditions. For example, it is desirable to extend our decomposition to handle nonlinear dynamics while preserving the high dimensionality and spatial information of the brain dynamics, and to compare these results with those obtained previously. We are currently working on extending our framework to nonlinear cases and will present the results elsewhere.

\subsection{An expression of the excess entropy production rate corresponding to the mode decomposition}

We discuss an expression of the excess entropy production rate $\sigma_t^\mathrm{ex}$ corresponding to the mode decomposition of the housekeeping entropy production rate. We can consider an expression of the excess entropy production rate by using a similar technique. First, we introduce the virtual dynamics driven by the excess part of the local mean velocity $\bm{\nu}_t^\mathrm{ex}(\bm{x}_s)$ instead of the housekeeping part $\bm{\nu}_t^\mathrm{hk}(\bm{x}_s)$,
\begin{align}
    d\bm{x}_s&=\bm{\nu}_t^\mathrm{ex}(\bm{x}_s)ds\nonumber\\
    &=\mathit{D}_t(\mathit{A}_t^\mathrm{ex}+\mathit{\Sigma}_t^{-1})\bm{x}_s ds+\mathit{D}_t\bm{b}^\mathrm{ex}_t-\mathit{D}_t\mathit{\Sigma}_t^{-1}\bm{\mu}_t.
    \label{vertualdynamicsex}
\end{align}
Since $\mathrm{eig} (\mathit{D}_t(\mathit{A}_t^\mathrm{ex} +\mathit{\Sigma}_t^{-1} ))= \mathrm{eig} (\sqrt{\mathit{D}_t} (\mathit{A}_t^\mathrm{ex} +\mathit{\Sigma}_t^{-1} )\sqrt{\mathit{D}_t})$, and $\sqrt{\mathit{D}_t} (\mathit{A}_t^\mathrm{ex} +\mathit{\Sigma}_t^{-1} )\sqrt{\mathit{D}_t}$ is the real symmetric matrix, all eigenvalues of $\mathit{D}_t(\mathit{A}_t^\mathrm{ex}+\mathit{\Sigma}_t^{-1})$ are real. Therefore, oscillatory behavior cannot be seen in the virtual dynamics [Eq.~(\ref{vertualdynamicsex})]. We  introduce the spectral decomposition $D_t(A_t^\mathrm{ex}+\mathit{\Sigma}_t^{-1}) = \sum_k \tilde{\lambda}_k \tilde{\mathsf{F}}_k$ with the real eigenvalue $\tilde{\lambda}_k$ and the projection matrix $\tilde{\mathsf{F}}_k= \tilde{\mathsf{P}} \bm{e}_k\bm{e}_k^\top \tilde{\mathsf{P}}^{-1}$ that satisfies $\sum_k \tilde{\mathsf{F}}_k = \mathit{I}$ and $\tilde{\mathsf{F}}_k \tilde{\mathsf{F}}_j = \delta_{kj} \tilde{\mathsf{F}}_k$, where $\tilde{\mathsf{P}}$ is a matrix of eigenvectors. The projection matrix $\tilde{\mathsf{F}}_k$ can be real  because the real matrix $\mathit{D}_t(\mathit{A}_t^\mathrm{ex} +\mathit{\Sigma}_t^{-1} )$ with all real eigenvalues has a real matrix of eigenvectors $\tilde{\mathsf{P}}$. Thus, $\tilde{\mathsf{F}}_k$ is chosen to be real. By using the spectral decomposition, we can rewrite the expression of the excess entropy production rate [Eq.~(\ref{eq:analytical-excess-epr})] as
\begin{align}
    \sigma_t^{\rm ex} =&(\mathit{A}^{\rm ex}_t \bm{\mu}_t + \bm{b}^{\rm ex}_t )^{\top} \mathit{D}_t (\mathit{A}^{\rm ex}_t \bm{\mu}_t + \bm{b}^{\rm ex}_t ) \nonumber \\
    &+\sum_{j,k} \tilde{\lambda}_j \tilde{\lambda}_k \trace \qty[   \mathit{D}^{-1}_t \tilde{\mathsf{F}}_j \mathit{\Sigma}_t \tilde{\mathsf{F}}_k^{\top}].
    \label{decompexcess}
\end{align}
Unlike in the case of the housekeeping entropy production rate, a contribution is made by two modes $\tilde{\lambda}_j \tilde{\lambda}_k \trace \qty[   \mathit{D}^{-1}_t \tilde{\mathsf{F}}_j \mathit{\Sigma}_t \tilde{\mathsf{F}}_k^{\top}]$, which is not necessarily nonnegative. Thus, this expression [Eq.~(\ref{decompexcess})] is not a decomposition of the excess entropy production rate into nonnegative contributions. Although this expression cannot be regarded as a mode decomposition, this expression [Eq.~(\ref{decompexcess})] might be useful in analyzing the relaxation in non-stationary dynamics, given that the time evolution of the virtual dynamics [Eq.~(\ref{vertualdynamicsex})] is governed by the factor $\exp(\tilde{\lambda}_k s)$ for each mode, and $1/|\tilde{\lambda}_k|$ implies the relaxation time for each mode if $\tilde{\lambda}_k$ is negative.

\section{Methods}\label{eq:appendix:data_analysis}
\subsection{Pre-processing}
We applied the main result to neural data. We used a publicly available 128-channel ECoG dataset recorded from a monkey ~\cite{Nagasaka2011-gk}. Data were recorded during awake eyes-open, awake eyes-closed, and ketamine-medetomidine-induced anesthesia. 

The data are pre-processed in the following steps. The data, originally sampled at 1000 Hz, were downsampled to 200 Hz using the \texttt{pop\_resample} function from EEGLAB~\cite{delorme2004eeglab}, with a cutoff frequency of 80 Hz and a transition bandwidth of 40 Hz. This was followed by a high pass filter applied at 2 Hz using the \texttt{pop\_eegfiltnew} function from EEGLAB~\cite{delorme2004eeglab} to remove trend components and increase data stationarity. Line noise at 50 Hz and 100 Hz was then removed using EEGLAB's \texttt{cleanline} function. To assess electrode quality, the standard deviation over time was calculated for each electrode. Electrodes with a standard deviation greater than three times the median standard deviation across all electrodes were considered poor and were subsequently removed. Finally, the data were rereferenced using a bipolar scheme. 
The downsampling and high-pass filter frequencies were chosen to avoid the artifacts described in Ref. \cite{barnett2011behaviour}, following the instructions in Ref. \cite{makoto_nd}.

\subsection{Model fitting}
The pre-processed data are fitted to the Langevin equation [Eq. (\ref{eq:Langevin_linear})].
The equation to be fit is
\begin{align}
    d\bm{x}_t=\mathit{D}\mathit{A}\bm{x}_tdt+\sqrt{2}\mathit{G}d\bm{B}_t \label{eq:Langevin_linear_methods}
\end{align} under the following assumptions. 
To plot the contribution of each electrode's oscillation to the housekeeping entropy production rate using Eq. (\ref{eq:EP_hk_decomposition_uniform}), we assume that the noise coefficient 
$\mathit{G}_t$ is a diagonal matrix.
In addition, the data were divided into 60-second time windows, and assumed to be stationary within each time window. Furthermore, within these time windows, both $\mathit{A}_t, \bm{b}, \mathit{D}_t$ and $\mathit{G}_t$ are assumed to be time invariant, and hence denoted as $\mathit{A}, \bm{b}, \mathit{D}$ and $\mathit{G}$, respectively. Furthermore, the mean over time was subtracted from the data, so that the mean of the data $\bm{\mu}_t$ is equal to the zero vector. This procedure makes $\bm{b}$ be the zero vector based on Eq. (\ref{eq:tv_mu}). 

The matrices $\mathit{D}\mathit{A}$ and $\mathit{D}$ were estimated by maximum likelihood estimation. 
The differential $d\bm{x}_t$ was approximated by the difference in the measured data for one time step.
Since the sampling rate of the preprocessed data was 200Hz, one time step corresponds to $dt=0.005$ seconds. The analytic expressions for the estimators of the matrix $\widehat{\mathit{DA}}$ and the diagonal elements of the matrix $2\widehat{\mathit{D}}$ are \begin{align}
    \widehat{\mathit{DA}}dt&=\qty(\sum_{t=1}^{N-1} d\bm{x}_t\bm{x}_t^\top)\qty(\sum_{t=1}^{N-1} \bm{x}_t\bm{x}_t^\top)^{-1}, \label{eq:estimation_DA}\\
    2\widehat{D}_{ii}dt&=\frac{1}{N-1}\sum_{t=1}^{N-1} \qty(dx_t^{(i)}-\qty(\widehat{\mathit{DA}})_{i:}\bm{x}_tdt)^2,
\end{align}
where $N$ is the number of time steps of the data, $dx_t^{(i)}$ is the $i$-th component of $d\bm{x}_t$, and $(\widehat{\mathit{DA}})_{i:}$ is the $i$-th row vector of the matrix $\widehat{\mathit{DA}}$. 
The symbol $\widehat{\ }$ represents an estimator and is used to distinguish it from the true value.
Note that $\widehat{\mathit{D}}$ is a diagonal matrix, since the noise coefficient $G$ is assumed to be a diagonal matrix. Using these estimators, the estimator of the matrix $\widehat{\mathit{A}}$ is given as
\begin{align}
    \widehat{\mathit{A}}=\widehat{\mathit{D}}^{-1}\widehat{\mathit{DA}}.
\end{align}
 
To satisfy the stationarity assumption under the estimated values of $\widehat{\mathit{DA}}$ and $\widehat{\mathit{D}}$ as described above, the covariance matrix $\widehat{\mathit{\Sigma}}$ was estimated as follows:
\begin{align}
    \widehat{\mathit{\Sigma}}=-\sum_{k,l} \frac{1}{\kappa_k+\overline{\kappa}_l}\mathsf{H}_k(2\widehat{D})\mathsf{H}_l^*,
\end{align}
where $\kappa_k$ is the $k$-th eigenvalue of $\widehat{\mathit{DA}}$ and $\mathsf{H}_k$ is the projection matrix that provides the spectoral decomposition of $\widehat{\mathit{DA}}$ as $\widehat{\mathit{DA}}=\sum_k \kappa_k \mathsf{H}_k$. This $\widehat{\mathit{\Sigma}}$ satisfies the condition of stationarity $\dot{\mathit{\Sigma}}=0$ in the Lyapunov equation (\ref{eq:Lyapunov}) because the projection matrix satisfies $\sum_k \mathsf{H}_k = I$ and $ \mathsf{H}_k \mathsf{H}_j = \delta_{kj} \mathsf{H}_k$. 

The excess part $\widehat{\mathit{A}^\mathrm{ex}}$ and the housekeeping part $\widehat{\mathit{A}^\mathrm{hk}}$ of $\widehat{\mathit{A}}$ are given as follows. Since we assumed $\widehat{\mathit{A}}$ and $\widehat{\mathit{\Sigma}}$ to be time-invariant, we omitted the subscript $t$ from $\widehat{\mathit{A}_t^\mathrm{ex}}$ and $\widehat{\mathit{A}_t^\mathrm{hk}}$. 
Since $\widehat{\mathit{A}^\mathrm{ex}}$ is a unique symmetry matrix satisfying (\ref{eq:Lyapunov_ex}), we can write \begin{align}
    \widehat{\mathit{A}^\mathrm{ex}}=-\widehat{\mathit{\Sigma}}^{-1} \label{eq:estimation_A_ex}, 
\end{align}
under the assumption that $\dot{\mathit{\Sigma}}=0$. Using $\widehat{\mathit{A}}$ in equation (\ref{eq:estimation_DA}) and $\widehat{\mathit{A}^\mathrm{ex}}$ in equation (\ref{eq:estimation_A_ex}), we get \begin{align}
    \widehat{\mathit{DA}^\mathrm{hk}}=\widehat{\mathit{DA}}-\widehat{\mathit{D}}\widehat{\mathit{A}^\mathrm{ex}}.
\end{align}
The spectral decomposition of $\widehat{\mathit{DA}^\mathrm{hk}}$ (Eq. \ref{eq:eigenvalue_decomposition}) allows us to apply our decomposition of the housekeeping entropy production rate into oscillatory components (Eq. \ref{eq:EP_hk_decomposition_uniform}) to data analysis. 

\subsection{Visualization of Results}
In Figs. 3 and 5, the contributions of each oscillatory mode $\sigma_t^\mathrm{hk, (k)}$ are binned into the delta band (0.5-4Hz), theta band (4-7Hz), alpha band (7-13Hz), and the frequency band below 0.5Hz, according to the conventional definition of frequency bands in the brain \cite{Clayton2018-js, Jensen2014-gz, Klimesch2012-un, Foxe2011-ve, Roux2014-vf, Sauseng2010-ah, Engel2010-om, Donoghue1998-ds}. The sums of the contributions from each frequency band are summarized as $\sigma_t^\mathrm{hk, <0.5}$, $\sigma_t^\mathrm{hk, Delta}$, $\sigma_t^\mathrm{hk, Theta}$, and $\sigma_t^\mathrm{hk, Alpha}$. They are defined as
\begin{align}
    \sigma_t^{\mathrm{hk, <0.5}}&=\sum_{k|0\leq |\chi_k|\leq 0.5}\sigma_t^{\mathrm{hk}(k)}\label{eq:def_sigma_05}\\
    \sigma_t^{\mathrm{hk, Delta}}&=\sum_{k|0.5\leq |\chi_k|\leq 4}\sigma_t^{\mathrm{hk}(k)}\label{eq:def_sigma_delta}\\
    \sigma_t^{\mathrm{hk, Theta}}&=\sum_{k|4\leq |\chi_k|\leq 7}\sigma_t^{\mathrm{hk}(k)}\label{eq:def_sigma_theta}\\
    \sigma_t^{\mathrm{hk, Alpha}}&=\sum_{k|7\leq |\chi_k|\leq 13}\sigma_t^{\mathrm{hk}(k)}.\label{eq:def_sigma_alpha}
\end{align}

In Fig. 4, the spatial distributions of the contributions to the entropy production rate from each frequency band $\bm{\sigma}_t^{\mathrm{hk, <0.5}}$, $\bm{\sigma}_t^{\mathrm{hk, Delta}}$, $\bm{\sigma}_t^{\mathrm{hk, Theta}}$, and $\bm{\sigma}_t^{\mathrm{hk, Alpha}}$ are shown. They are defined as
\begin{align}
    \bm{\sigma}_t^{\mathrm{hk, <0.5}} &= (\sigma_t^{\mathrm{hk, <0.5}, (i=1)}, \cdots, \sigma_t^{\mathrm{hk, <0.5}, (i=d)}) \label{eq:def_sigma_05_vec}\\
    \bm{\sigma}_t^{\mathrm{hk, Delta}} &= (\sigma_t^{\mathrm{hk, Delta}, (i=1)}, \cdots, \sigma_t^{\mathrm{hk, Delta}, (i=d)}) \label{eq:def_sigma_delta_vec}\\    
    \bm{\sigma}_t^{\mathrm{hk, Theta}} &= (\sigma_t^{\mathrm{hk, Theta}, (i=1)}, \cdots, \sigma_t^{\mathrm{hk, Theta}, (i=d)}) \label{eq:def_sigma_theta_vec}\\
    \bm{\sigma}_t^{\mathrm{hk, Alpha}} &= (\sigma_t^{\mathrm{hk, Alpha}, (i=1)}, \cdots, \sigma_t^{\mathrm{hk, Alpha}, (i=d)}), \label{eq:def_sigma_alpha_vec}
\end{align}
where
\begin{align}
    \sigma_t^{\mathrm{hk, <0.5}, (i)}&=\sum_{k|0\leq |\chi_k|\leq 0.5}\sigma_t^{\mathrm{hk}(k,i)}\label{eq:def_sigma_05_i}\\
    \sigma_t^{\mathrm{hk, Delta}, (i)}&=\sum_{k|0.5\leq |\chi_k|\leq 4}\sigma_t^{\mathrm{hk}(k,i)}\label{eq:def_sigma_delta_i}\\
    \sigma_t^{\mathrm{hk, Theta}, (i)}&=\sum_{k|4\leq |\chi_k|\leq 7}\sigma_t^{\mathrm{hk}(k,i)}\label{eq:def_sigma_theta_i}\\
    \sigma_t^{\mathrm{hk, Alpha}, (i)}&=\sum_{k|7\leq |\chi_k|\leq 13}\sigma_t^{\mathrm{hk}(k,i)}.\label{eq:def_sigma_alpha_i}
\end{align}

\begin{acknowledgments}
D.S., S.I. and M.O. thank Andreas Dechant for fruitful discussions on a mode decomposition. 
D.S. and M.O. thank Daiki Kiyooka for helpful discussions and S.I. thanks Ryuna Nagayama, Kohei Yoshimura and Artemy Kolchinsky for fruitful discussions on a geometric decomposition. 
D.S. is supported by JSPS KAKENHI Grants No. 23KJ0799.
S.~I. is supported by JSPS KAKENHI Grants No. 19H05796,
No.~21H01560, No.~22H01141, and No.~23H00467, JST ERATO Grant No.~JPMJER2302, and UTEC-UTokyo
FSI Research Grant Program. 
M.O. is supported by JST Moonshot R\&D Grant Number JPMJMS2012,  JST CREST Grant Number JPMJCR1864, and Japan Promotion Science, Grant-in-Aid for Transformative Research Areas Grant Numbers 20H05712, 23H04834. 
\end{acknowledgments}

\appendix

\section{Detailed derivation of Eq.~(\ref{eq:kldivergence})} \label{eq:appendixA}
To derive Eq.~(\ref{eq:kldivergence}), we start with the following expression,
\begin{align}
&\mathrm{D_{KL}}[\mathbb{P}_{\rm
F}\|\mathbb{P}_{\rm B}] \nonumber \\
=& \int d\boldsymbol{x}_t d\boldsymbol{x}_{t+dt} \mathbb{P}_{\rm
F} (\boldsymbol{x}_{t+dt} ,\boldsymbol{x}_{t})  \ln \frac{\mathbb{T} (\boldsymbol{x}_{t+dt} |\boldsymbol{x}_{t}) p_t (\boldsymbol{x}_t)}{\mathbb{T} (\boldsymbol{x}_{t} |\boldsymbol{x}_{t+dt}) p_{t+dt} (\boldsymbol{x}_{t+dt})} \nonumber \\
=& \int d\boldsymbol{x}_t d\boldsymbol{x}_{t+dt} \mathbb{P}_{\rm
F} (\boldsymbol{x}_{t+dt} ,\boldsymbol{x}_{t}) \ln \frac{\mathbb{T} (\boldsymbol{x}_{t+dt} |\boldsymbol{x}_{t})p_{t} (\boldsymbol{x}_{t}) }{\mathbb{T} (\boldsymbol{x}_{t} |\boldsymbol{x}_{t+dt})p_{t} (\boldsymbol{x}_{t+dt})} \nonumber \\
&+\int d\boldsymbol{x}_{t+dt} p_{t+dt}(\boldsymbol{x}_{t+dt}) \ln \frac{p_t (\boldsymbol{x}_{t+dt}) }{p_{t+dt}(\boldsymbol{x}_{t+dt})},
\end{align}
where we used the fact $ \int d\boldsymbol{x}_t  \mathbb{P}_{\rm
F} (\boldsymbol{x}_{t+dt} ,\boldsymbol{x}_{t})  = p_{t+dt}(\boldsymbol{x}_{t+dt}) $.
Here, the last term 
\begin{align}
\int d\boldsymbol{x}_{t+dt} p_{t+dt}(\boldsymbol{x}_{t+dt}) \ln \frac{p_t (\boldsymbol{x}_{t+dt}) }{p_{t+dt}(\boldsymbol{x}_{t+dt})} = O(dt^2),
\end{align}
is negligible. By using the  transition probability Eq.~(\ref{eq:condprob}), we can calculate the quantity $\ln  [\mathbb{T} (\boldsymbol{x}_{t+dt} |\boldsymbol{x}_{t})p_{t} (\boldsymbol{x}_{t}) ]/[ \mathbb{T} (\boldsymbol{x}_{t} |\boldsymbol{x}_{t+dt})p_{t} (\boldsymbol{x}_{t+dt})]$ as 
\begin{align}
&\ln \frac{\mathbb{T} (\boldsymbol{x}_{t+dt} |\boldsymbol{x}_{t})p_{t} (\boldsymbol{x}_{t}) }{\mathbb{T} (\boldsymbol{x}_{t} |\boldsymbol{x}_{t+dt})p_{t} (\boldsymbol{x}_{t+dt})} \nonumber\\ 
=& \frac{(\dot{\boldsymbol{x}}_t )^{\top} (\boldsymbol{f}_t (\boldsymbol{x}_t)+\boldsymbol{f}_t (\boldsymbol{x}_{t+dt}) ) dt}{2} \nonumber \\
&-  \frac{(\dot{\boldsymbol{x}}_t )^{\top} [\nabla \ln p_t(\boldsymbol{x}_t) +\nabla \ln p_{t}(\boldsymbol{x}_{t+dt}) ]}{2} + C \nonumber \\
=&  D^{-1}_t \boldsymbol{\nu}_t (\boldsymbol{x}_t) \circ \dot{\boldsymbol{x}}_t  dt + C,
\end{align}
where $\circ$ stands for Stratonovich integral, which is defined as $\boldsymbol{y}(\boldsymbol{x}_t)\circ \dot{\boldsymbol{x}}_t = (\boldsymbol{x}_{t+dt} - \boldsymbol{x}_t)^{\top}[\boldsymbol{y}(\boldsymbol{x}_t) +  \boldsymbol{y}(\boldsymbol{x}_{t+dt})]/(2dt)$ for any vector $\boldsymbol{y}(\boldsymbol{x}_t)$, and $C$ is the negligible term that satisfies $\int d\boldsymbol{x}_t d\boldsymbol{x}_{t+dt} \mathbb{P}_{\rm
F} (\boldsymbol{x}_{t+dt} ,\boldsymbol{x}_{t}) C =O(dt^2)$. By neglecting $O(dt^2)$, the Kullback--Leibler divergence is calculated as
\begin{align}
&\mathrm{D_{KL}}[\mathbb{P}_{\rm
F}\|\mathbb{P}_{\rm B}] \nonumber \\
&=\int d\boldsymbol{x}_t d\boldsymbol{x}_{t+dt} \mathbb{T}(\boldsymbol{x}_{t+dt}| \boldsymbol{x}_t)  p_t(\boldsymbol{x}_t) D^{-1}_t \boldsymbol{\nu}_t(\boldsymbol{x}_t) \circ \dot{\boldsymbol{x}}_t dt.
\label{eq:calculationkl}
\end{align}

To obtain Eq.~(\ref{eq:kldivergence}) from the expression in Eq.~(\ref{eq:calculationkl}), we consider the following formula 
\begin{align}
&\int d\boldsymbol{x}_t d\boldsymbol{x}_{t+dt} \mathbb{T}(\boldsymbol{x}_{t+dt}| \boldsymbol{x}_t)  p_t(\boldsymbol{x}_t) \boldsymbol{y}(\boldsymbol{x}_t) \circ  \dot{\boldsymbol{x}}_t dt \nonumber \\
&= \int d\boldsymbol{x}_t p_t(\boldsymbol{x}_t)  (\boldsymbol{y}(\boldsymbol{x}_t) )^{\top}\boldsymbol{\nu}_t(\boldsymbol{x}_t)   dt, \label{eq:formula}
\end{align}
for any vector $\boldsymbol{y}(\boldsymbol{x}_t)$.
To prove it, we start with the following Gaussian integral,
\begin{align}
&\int d\boldsymbol{x}_{t+dt} \mathbb{T}(\boldsymbol{x}_{t+dt}| \boldsymbol{x}_t)   \boldsymbol{y}(\boldsymbol{x}_t) \circ d\boldsymbol{x}_t dt \nonumber \\
=& \int d\boldsymbol{x}_t \mathbb{T}(\boldsymbol{x}_{t+dt}| \boldsymbol{x}_t)  \left[\frac{\boldsymbol{y}(\boldsymbol{x}_t) + \boldsymbol{y}(\boldsymbol{x}_{t+dt})}{2} \right]^{\top} (\boldsymbol{x}_{t+dt} - \boldsymbol{x}_t) \nonumber \\
=& \int d\boldsymbol{x}_t \mathbb{T}(\boldsymbol{x}_{t+dt}| \boldsymbol{x}_t) [\boldsymbol{y}(\boldsymbol{x}_t) ]^{\top}(\boldsymbol{x}_{t+dt} - \boldsymbol{x}_t) \nonumber \\
&+\sum_{i,j}\int d\boldsymbol{x}_t \mathbb{T}(\boldsymbol{x}_{t+dt}| \boldsymbol{x}_t) \nonumber\\
&\: \: \: \times \frac{[(\nabla)_j (\boldsymbol{y}(\boldsymbol{x}_t))_i](\boldsymbol{x}_{t+dt} - \boldsymbol{x}_t)_i (\boldsymbol{x}_{t+dt} - \boldsymbol{x}_t)_j}{2} \nonumber \\
=& [\boldsymbol{y}(\boldsymbol{x}_t) ]^{\top}D_t \boldsymbol{f}_t (\boldsymbol{x}_t) dt + \sum_{i,j} [(\nabla)_j (\boldsymbol{y}(\boldsymbol{x}_t))_i] (D_t)_{ij} dt
\end{align}
where we neglect the higher order term $O(dt^2)$. Thus, we obtain the formula as follows, 
\begin{align}
&\int d\boldsymbol{x}_t d\boldsymbol{x}_{t+dt} \mathbb{T}(\boldsymbol{x}_{t+dt}| \boldsymbol{x}_t)  p_t(\boldsymbol{x}_t) \boldsymbol{y}(\boldsymbol{x}_t) \circ d\boldsymbol{x}_t dt \nonumber \\
=& \int d\boldsymbol{x}_t  p_t(\boldsymbol{x}_t) [\boldsymbol{y}(\boldsymbol{x}_t) ]^{\top}D_t \boldsymbol{f}_t (\boldsymbol{x}_t) dt \nonumber \\
&+ \sum_{i,j}\int d\boldsymbol{x}_t  p_t(\boldsymbol{x}_t)  [(\nabla)_j (\boldsymbol{y}(\boldsymbol{x}_t))_i] (D_t)_{ij} dt \nonumber\\
=& \int d\boldsymbol{x}_t  p_t(\boldsymbol{x}_t) [\boldsymbol{y}(\boldsymbol{x}_t) ]^{\top}D_t \boldsymbol{f}_t (\boldsymbol{x}_t) dt \nonumber \\
&-\int d\boldsymbol{x}_t   p_t  (\boldsymbol{x}_t) [\boldsymbol{y}(\boldsymbol{x}_t)]^{\top} D_t\nabla \ln p_t (\boldsymbol{x}_t) dt \nonumber\\
=&\int d\boldsymbol{x}_t p_t(\boldsymbol{x}_t)  (\boldsymbol{y}(\boldsymbol{x}_t) )^{\top}\boldsymbol{\nu}_t(\boldsymbol{x}_t)   dt,
\end{align}
where we used the partial integration with the assumption that $p_t (\boldsymbol{x}_t) \to 0$ in the limit $\|\boldsymbol{x}_t\| \to \infty$. 

By using the formula Eq.~(\ref{eq:formula}) for $\boldsymbol{y}(\boldsymbol{x}_t) = D_t^{-1} \boldsymbol{\nu}_t(\boldsymbol{x}_t)$,
we can obtain Eq.~(\ref{eq:kldivergence}) from the expression in Eq.~(\ref{eq:calculationkl}),
\begin{align}
\mathrm{D_{KL}}[\mathbb{P}_{\rm
F}\|\mathbb{P}_{\rm B}] &= \int d\boldsymbol{x}_t p_t(\boldsymbol{x}_t)  (D^{-1}_t \boldsymbol{\nu}_t(\boldsymbol{x}_t))^{\top}\boldsymbol{\nu}_t(\boldsymbol{x}_t)   dt \nonumber\\
&= \sigma_t dt.
\end{align}

\section{Details of the geometric decomposition}\label{appendix:excess_wasserstein}
First, we explain the relation between the excess entropy production rate and optimal transport theory~\cite{villani2009optimal}.
When the noise covariance $\mathit{D}_t$ is a scalar matrix and written as $\mathit{D}_t=TI$ using the identity matrix $I$, the excess entropy production rate $\sigma^\mathrm{ex}_t$ can be written using the $L^2$-Wasserstein distance $\mathcal{W}_2$~\cite{villani2009optimal} in the optimal transport theory~\cite{Dechant2022-gt, ito2023geometric}:
\begin{align}
\sigma^\mathrm{ex}_t&=\min_{\bm{\nu}_t|\partial {p}_t/\partial t=-\nabla\cdot[\bm{\nu}_t p_t]}\frac{1}{T}\langle \|\bm{\nu}_t \|^2 \rangle_t \nonumber \\
&=\frac{1}{T}\langle \|\nabla \phi_t\|^2 \rangle_t  \nonumber \\
    &=\frac{1}{T} \lim_{\Delta t \to 0}\frac{\mathcal{W}_2\qty(p_t, p_{t+\Delta t})^2}{(\Delta t)^2}.
\end{align}
This expression is a consequence of the Benamou-Brenier formula~\cite{benamou2000computational} for an infinitesimal time interval in optimal transport theory~\cite{Nakazato2021-do,Dechant2022-gt,ito2023geometric}. This geometric interpretation of the excess entropy production rate leads to the thermodynamic speed limit, which is a fundamental thermodynamic limit on the speed~\cite{aurell2012refined,Nakazato2021-do,ito2023geometric}.  

Next, we compare the geometric decomposition with a different decomposition introduced by Hatano and Sasa~\cite{hatano2001steady}. The decomposition introduced by Hatano and Sasa uses the steady state velocity field $\boldsymbol{\nu}_t^{\rm st}$ which satisfies $0=-\nabla \cdot \qty[\bm{\nu}_t^\mathrm{st}(\bm{x})p^{\rm st}_t(\bm{x})]$ with the steady state distribution $p^{\rm st}_t (\boldsymbol{x})$. The housekeeping entropy production rate introduced by Hatano and Sasa is defined as $\sigma^{\rm hk;HS}_t = \langle (\bm{\nu}_t^\mathrm{st})^{\top} D_t^{-1} \bm{\nu}_t^\mathrm{st} \rangle_t$. Whereas $\sigma^{\rm hk;HS}_t$ depends on the steady-state distribution $p^{\rm st}_t(\bm{x})$, the housekeeping entropy production rate in a geometric decomposition $\sigma^{\rm hk}_t$ only depends on the current distribution $p_t(\bm{x})$. In general, $\sigma^{\rm hk;HS}_t$ is not equivalent to  $\sigma^{\rm hk}_t$, and the inequality $\sigma^{\rm hk;HS}_t \geq \sigma^{\rm hk}_t$ holds~\cite{dechant2022geometric}. In the steady state, the excess entropy production rate vanishes and the two housekeeping entropy production rates are equivalent to the entropy production rate $\sigma_t=\sigma^{\rm hk;HS}_t = \sigma^{\rm hk}_t$.

\section{Geometric expressions of the geometric decomposition of the entropy production rates} \label{appendix:HS-innerproduct} 
To understand the analytical expressions of the entropy production rates [Eqs.~(\ref{eq:analytical-epr}), (\ref{eq:analytical-excess-epr}) and (\ref{eq:analytical-housekeeping-epr})] from the viewpoint of geometry, we can introduce the Hilbert-Schmidt inner product. For any real-valued matrices $\mathit{Y}$ and $\mathit{Z}$, the Hilbert-Schmidt inner product is defined as $\langle \mathit{Y}, \mathit{Z} \rangle_{\rm HS} = \trace[\mathit{Y}^{\top} \mathit{Z}] =\trace[\mathit{Z}^{\top} \mathit{Y}]$. For any vectors $\bm{y}$ and $\bm{z}$, the Hilbert-Schmidt inner product is regarded as the conventional inner product $\langle \bm{y} , \bm{z} \rangle_{\rm HS} = \bm{y}^{\top} \bm{z} = \bm{z}^{\top} \bm{y}$. The Hilbert-Schmidt norm is introduced as $\|\mathit{Y}\|_{\rm HS} = \sqrt{\langle \mathit{Y} , \mathit{Y} \rangle_{\rm HS} }$ and the $L^2$-norm is introduced as $\|\bm{y} \| = \sqrt{\langle \bm{y} , \bm{y} \rangle_{\rm HS} }$. Using the Hilbert-Schmidt norm and the $L^2$-norm, the entropy production rates are rewritten as
\begin{align}
    \sigma_t =& \| \mathit{G}_t^{\top} (\mathit{A}_t \bm{\mu}_t + \bm{b}_t ) \|^2  +  \| \mathit{G}_t^{\top}  (\mathit{A}_t + \mathit{\Sigma}_t^{-1} ) \sqrt{\mathit{\Sigma}_t}  \|^2_{\rm HS}, \\
    \sigma_t^{\rm ex} =&\| \mathit{G}_t^{\top} (\mathit{A}^{\rm ex}_t \bm{\mu}_t + \bm{b}^{\rm ex}_t ) \|^2  +  \| \mathit{G}_t^{\top}  (\mathit{A}^{\rm ex}_t + \mathit{\Sigma}_t^{-1} ) \sqrt{\mathit{\Sigma}_t}  \|^2_{\rm HS},  \\
    \sigma^\mathrm{hk}_t =&\| \mathit{G}_t^{\top} \mathit{A}^{\rm hk}_t \sqrt{\mathit{\Sigma}_t}]  \|^2_{\rm HS}. 
\end{align}
These geometric interpretations using the Hilbert-Schmidt norm and the $L^2$-norm provide the non-negativity of the entropy production rates, $\sigma_t \geq 0$, $\sigma_t^{\rm ex} \geq 0$ and $\sigma_t^{\rm hk} \geq 0$.

Based on these geometric interpretations, we can check the validity of the geometric decomposition $\sigma_t= \sigma^{\rm ex}_t+ \sigma_t^{\rm hk}$ as follows. From Eq.~(\ref{eq:tv_mu_hk}), we obtain an equivalence of the two $L^2$-norms,  
\begin{align}
    \| \mathit{G}_t^{\top} (\mathit{A}_t \bm{\mu}_t + \bm{b}_t ) \|^2   =&\| \mathit{G}_t^{\top} (\mathit{A}^{\rm ex}_t \bm{\mu}_t + \bm{b}^{\rm ex}_t ) \|^2. \label{eq:equivalence}
\end{align}
We can also obtain the orthogonality
\begin{align}
    &\langle \mathit{G}_t^{\top}  (\mathit{A}^{\rm ex}_t + \mathit{\Sigma}_t^{-1} ) \sqrt{\mathit{\Sigma}_t}  , \mathit{G}_t^{\top} \mathit{A}^{\rm hk}_t \sqrt{\mathit{\Sigma}_t} \rangle_{\rm HS} = 0, \label{eq:orthogonality}
\end{align}
because we can calculate this Hilbert-Schmidt inner product as
\begin{align}
    &\langle \mathit{G}_t^{\top}  (\mathit{A}^{\rm ex}_t + \mathit{\Sigma}_t^{-1} ) \sqrt{\mathit{\Sigma}_t}  , \mathit{G}_t^{\top} \mathit{A}^{\rm hk}_t \sqrt{\mathit{\Sigma}_t} \rangle_{\rm HS}   \nonumber \\
    &= \trace [ (\mathit{A}^{\rm ex}_t + \mathit{\Sigma}_t^{-1} )^{\top}  \mathit{D}_t \mathit{A}^{\rm hk}_t \mathit{\Sigma}_t]
 \nonumber\\
 &= - \trace [  \mathit{\Sigma}_t (\mathit{A}^{\rm hk}_t)^{\top}  \mathit{D}_t (\mathit{A}^{\rm ex}_t + \mathit{\Sigma}_t^{-1} )  ] \nonumber\\
 &= -\langle  \mathit{G}_t^{\top} \mathit{A}^{\rm hk}_t \sqrt{\mathit{\Sigma}_t}, \mathit{G}_t^{\top}  (\mathit{A}^{\rm ex}_t + \mathit{\Sigma}_t^{-1} ) \sqrt{\mathit{\Sigma}_t}  \rangle_{\rm HS} \nonumber\\
 &= -\langle \mathit{G}_t^{\top}  (\mathit{A}^{\rm ex}_t + \mathit{\Sigma}_t^{-1} ) \sqrt{\mathit{\Sigma}_t}  , \mathit{G}_t^{\top} \mathit{A}^{\rm hk}_t \sqrt{\mathit{\Sigma}_t} \rangle_{\rm HS} ,
\end{align}
where we used Eq.~(\ref{eq:lyapunov_hk}), $(\mathit{A}^{\rm ex}_t)^{\top}= \mathit{A}^{\rm ex}_t$,  $\mathit{D}_t=\mathit{G}_t\mathit{G}_t^\top$, $(\mathit{\Sigma}_t^{-1})^{\top}= \mathit{\Sigma}_t^{-1}$,  $\mathit{D}_t^{\top}= \mathit{D}_t$ and the cyclic property of the trace. This orthogonality [Eq.~(\ref{eq:orthogonality})] provides a decomposition
\begin{align}
& \| \mathit{G}_t^{\top}  (\mathit{A}_t + \mathit{\Sigma}_t^{-1} ) \sqrt{\mathit{\Sigma}_t}  \|^2_{\rm HS} \nonumber\\
&=\| \mathit{G}_t^{\top}  (\mathit{A}^{\rm ex}_t + \mathit{\Sigma}_t^{-1} ) \sqrt{\mathit{\Sigma}_t} \|^2_{\rm HS} +\| \mathit{G}_t^{\top} \mathit{A}^{\rm hk}_t \sqrt{\mathit{\Sigma}_t}]  \|^2_{\rm HS},
\label{eq:pythagorean}
\end{align}
where we used the Pythagorean theorem $\langle Y+Z, Y+Z \rangle_{\rm HS} = \langle Y, Y \rangle_{\rm HS} + \langle Z, Z \rangle_{\rm HS}$ with $\langle Y, Z\rangle_{\rm HS}=0$ by adopting $Y=\mathit{G}_t^{\top}  (\mathit{A}^{\rm ex}_t + \mathit{\Sigma}_t^{-1} ) \sqrt{\mathit{\Sigma}_t}$ and $Z=\mathit{G}_t^{\top} \mathit{A}^{\rm hk}_t \sqrt{\mathit{\Sigma}_t}$. By combining two geometric relations (\ref{eq:equivalence}) and (\ref{eq:pythagorean}), we obtain the geometric decomposition $\sigma_t= \sigma^{\rm ex}_t+ \sigma_t^{\rm hk}$.

\section{Invariance of our decomposition with respect to linear transformation} \label{appendix_invariance_for_linear_transform}
We discuss the invariance of the housekeeping entropy production rate under the linear transformation in order to justify the use of the center of mass coordinate and the relative coordinate in Eq.~(\ref{eq:langevin_centor_relative}).
For the original linear Langevin equation (\ref{eq:Langevin_linear}), we consider the linear transformation \begin{align}
    \bm{y}_t=R\bm{x}_t,
\end{align}
where $R$ is a real regular matrix.
The Langevin equation is rewritten as
\begin{align}
d\bm{y}_t=D_t'(A_t'\bm{y}_t+\bm{b}_t')dt+\sqrt{2}G_t'd\bm{B}_t,
\end{align}
where $\mathit{\Sigma}_t'$ is the covariance matrix for $\bm{y}_t$ and
\begin{align}
    G_t'&=R G_t, \\
    D_t'&=G_t'G_t'^\top =RD_tR^\top,\\
    A_t'&=(R^{-1})^\top A_tR^{-1}, \\
    \bm{b}_t'&=(R^{-1})^\top  \bm{b}_t.
\end{align}  
The mean values and the variance-covariance matrix for $\bm{y}_t$ is also rewritten as 
\begin{align}
\mathit{\Sigma}_t'&=R\mathit{\Sigma}_t R^\top, \\
    \bm{\mu}_t' &=R \bm{\mu}_t.
\end{align}

The decompositions of $A_t'$ and $\bm{b}_t'$ for $\bm{y}_t$ into excess and housekeeping parts is written using $A_t^\mathrm{ex}$, $A_t^\mathrm{hk}$, $\bm{b}_t^\mathrm{ex}$ and $\bm{b}_t^\mathrm{hk}$ for $\bm{x}_t$ as
\begin{align}
     (A_t') ^\mathrm{ex}&=(R^{-1})^\top A_t^\mathrm{ex}R^{-1}, \hspace{5mm} (A_t')^\mathrm{hk}=(R^{-1})^\top A_t^\mathrm{hk}R^{-1} \nonumber\\
    (\bm{b}_t')^\mathrm{ex}&=(R^{-1})^\top \bm{b}_t^\mathrm{ex}, \hspace{5mm} (\bm{b}_t')^\mathrm{hk}=(R^{-1})^\top \bm{b}_t^\mathrm{hk},
\end{align}
This is verified by the fact that $(A_t') ^\mathrm{ex}$ satisfies the condition of symmetric matrix, and by the fact that these satisfy the conditions corresponding to Eqs. (\ref{eq:tv_mu_ex}), (\ref{eq:Lyapunov_ex}), (\ref{eq:tv_mu_hk}) and (\ref{eq:lyapunov_hk}): 
\begin{align}
    D_t'(A_t') ^\mathrm{ex}\bm{\mu}_t'+D_t' (\bm{b}_t')^\mathrm{ex}&=\dot{\bm{\mu}}_t',\\
    D_t'(A_t') ^\mathrm{ex}\mathit{\Sigma}_t'+\mathit{\Sigma}_t' (D_t'(A_t') ^\mathrm{ex})^\top +2D' &=\dot{\mathit{\Sigma}}_t' ,\\
    D_t'(A_t') ^\mathrm{hk}\bm{\mu}_t'+ D_t'(\bm{b}_t')^\mathrm{hk} &=\bm{0},\\
    D_t'(A_t') ^\mathrm{hk}\mathit{\Sigma}'+\mathit{\Sigma}'(D_t'(A_t') ^\mathrm{hk} )^\top &=O.
\end{align}
Thus, we can show that the entropy production rate and the excess and housekeeping entropy productions [Eqs.~(\ref{eq:analytical-epr}), (\ref{eq:analytical-excess-epr}) and (\ref{eq:analytical-housekeeping-epr})] are invariant under the linear transformation
as follows,
\begin{align}
    \sigma_t =&(\mathit{A}_t \bm{\mu}_t + \bm{b}_t )^{\top} \mathit{D}_t (\mathit{A}_t \bm{\mu}_t + \bm{b}_t ) \nonumber \\
    &+ \trace \qty[ (\mathit{A}_t + \mathit{\Sigma}_t^{-1} )^{\top} \mathit{D}_t  (\mathit{A}_t + \mathit{\Sigma}_t^{-1} )\mathit{\Sigma}_t]\nonumber \\
    =&(\mathit{A}_t' \bm{\mu}_t' + \bm{b}_t' )^{\top} \mathit{D}_t '(\mathit{A}_t' \bm{\mu}_t' + \bm{b}_t ') \nonumber \\
    &+ \trace \qty[ (\mathit{A}_t '+ (\mathit{\Sigma}_t')^{-1})^{\top} \mathit{D}_t'  (\mathit{A}_t' + (\mathit{\Sigma}_t')^{-1} )\mathit{\Sigma}_t']\\
    \sigma_t^{\rm ex} =&(\mathit{A}^{\rm ex}_t \bm{\mu}_t + \bm{b}^{\rm ex}_t )^{\top} \mathit{D}_t (\mathit{A}^{\rm ex}_t \bm{\mu}_t + \bm{b}^{\rm ex}_t ) \nonumber \\
    &+ \trace \qty[ (\mathit{A}^{\rm ex}_t + \mathit{\Sigma}_t^{-1} )^{\top} \mathit{D}_t  (\mathit{A}^{\rm ex}_t + \mathit{\Sigma}_t^{-1} )\mathit{\Sigma}_t] \nonumber\\
    =&((\mathit{A}'_t)^{\rm ex} \bm{\mu}'_t + (\bm{b}'_t)^{\rm ex} )^{\top} \mathit{D}'_t ((\mathit{A}'_t)^{\rm ex} \bm{\mu}'_t + (\bm{b}'_t)^{\rm ex}  ) \nonumber \\
    &+ \trace \qty[ ((\mathit{A}'_t)^{\rm ex} + (\mathit{\Sigma}_t')^{-1} )^{\top} \mathit{D}'_t  ((\mathit{A}'_t)^{\rm ex} + (\mathit{\Sigma}_t')^{-1})\mathit{\Sigma}'_t]\\
    \sigma^\mathrm{hk}_t =& \trace \qty[ (\mathit{A}_t^{\rm hk})^{\top} \mathit{D}_t  \mathit{A}_t^{\rm hk} \mathit{\Sigma}_t] \nonumber\\
    =& \trace \qty[ ((\mathit{A}'_t)^{\rm hk})^{\top} \mathit{D}'_t  (\mathit{A}'_t)^{\rm hk} \mathit{\Sigma}'_t],\label{eq:analytical-housekeeping-epr_invariance}
\end{align}
where we use the cyclic property of the trace.

Additionally, we note the invariance of the housekeeping entropy production rate under the linear transformation in terms of the mode decomposition. Consider the spectral decomposition of $D_t'(A_t')^\mathrm{hk} = R D_t A_t^\mathrm{hk} R^{-1}$. By using the spectral decomposition of $D_t A_t^\mathrm{hk}$ (\ref{eq:eigenvalue_decomposition}), the spectral decomposition $D_t'(A_t')^\mathrm{hk}$ is obtained as \begin{align}
    D_t'(A_t')^\mathrm{hk}=\sum_k \lambda_k R\mathsf{F}_kR^{-1}=\sum_k \lambda_k \mathsf{F}'_k,
\end{align}
where we put $R\mathsf{F}_kR^{-1}=\mathsf{F}_k'$. We can confirm that $\mathsf{F}_k'$ is also a projection matrix because $\sum_k \mathsf{F}'_k=I$ and $\mathsf{F}'_k\mathsf{F}'_l=\delta_{kl}\mathsf{F}'_k$ are satisfied. Thus, the eigenvalues of $D_t'(A_t')^\mathrm{hk}$ are the same as the eigenvalues of $D_t(A_t)^\mathrm{hk}$. We can check the following invariance
\begin{align}
 &    \trace \qty(\mathit{G}_t^{-1}\mathsf{F}_k\mathit{\Sigma}_t\mathsf{F}_k ^* (\mathit{G}_t^{-1})^{\top} ) \nonumber \\
&= \trace \qty( (\mathit{G}_t')^{-1}\mathsf{F}_k' \mathit{\Sigma}_t' (\mathsf{F}_k')^* ((\mathit{G}_t')^{-1})^{\top} ),
\end{align}
and in this way we obtain the invariance of the housekeeping entropy production rate under the linear transformation in terms of the mode decomposition,
\begin{align}
    \sigma^\mathrm{hk}_t
    &=\sum_{k} |\lambda_k|^2 \trace \qty(\mathit{G}_t^{-1}\mathsf{F}_k\mathit{\Sigma}_t\mathsf{F}_k ^* (\mathit{G}_t^{-1})^{\top} ) \nonumber \\
    &= \sum_{k} |\lambda_k|^2 \trace \qty( (\mathit{G}_t')^{-1}\mathsf{F}_k' \mathit{\Sigma}_t' (\mathsf{F}_k')^* ((\mathit{G}_t')^{-1})^{\top} ).
\end{align}

\section{Robustness of temporal stability of the decomposition across all individual monkeys}
The temporal stability of the decomposition was robust across all individual monkeys. 
Similarly to Fig. \ref{fig:time_variatiom}b-e, the sums of the contributions from each frequency band $ \sigma_t^\mathrm{hk, <0.5}$, $ \sigma_t^\mathrm{hk, Delta}$, $ \sigma_t^\mathrm{hk, Theta}$, and $ \sigma_t^\mathrm{hk, Alpha}$ calculated from each time window are shown in the dot plots, and their averages across time windows are shown in the bar plots (Fig. \ref{fig:time_variation_all-monkey}). 
For all individual monkeys, we observed that the contributions of the delta band $\sigma_t^\mathrm{hk, Delta}$ were larger in the anesthetized condition than in the awake conditions, and that the contributions of the theta band $\sigma_t^\mathrm{hk, Theta}$ were larger in the awake conditions than in the anesthetized condition. The dot plots in Fig. \ref{fig:time_variation_all-monkey} show that the time variations of $\sigma_t^\mathrm{hk, Delta}$ and $ \sigma_t^\mathrm{hk, Theta}$ were smaller than these differences across conditions.

\setcounter{figure}{0}
\renewcommand{\thefigure}{S\arabic{figure}}

\begin{figure*}
\centering
\includegraphics[width=\linewidth]{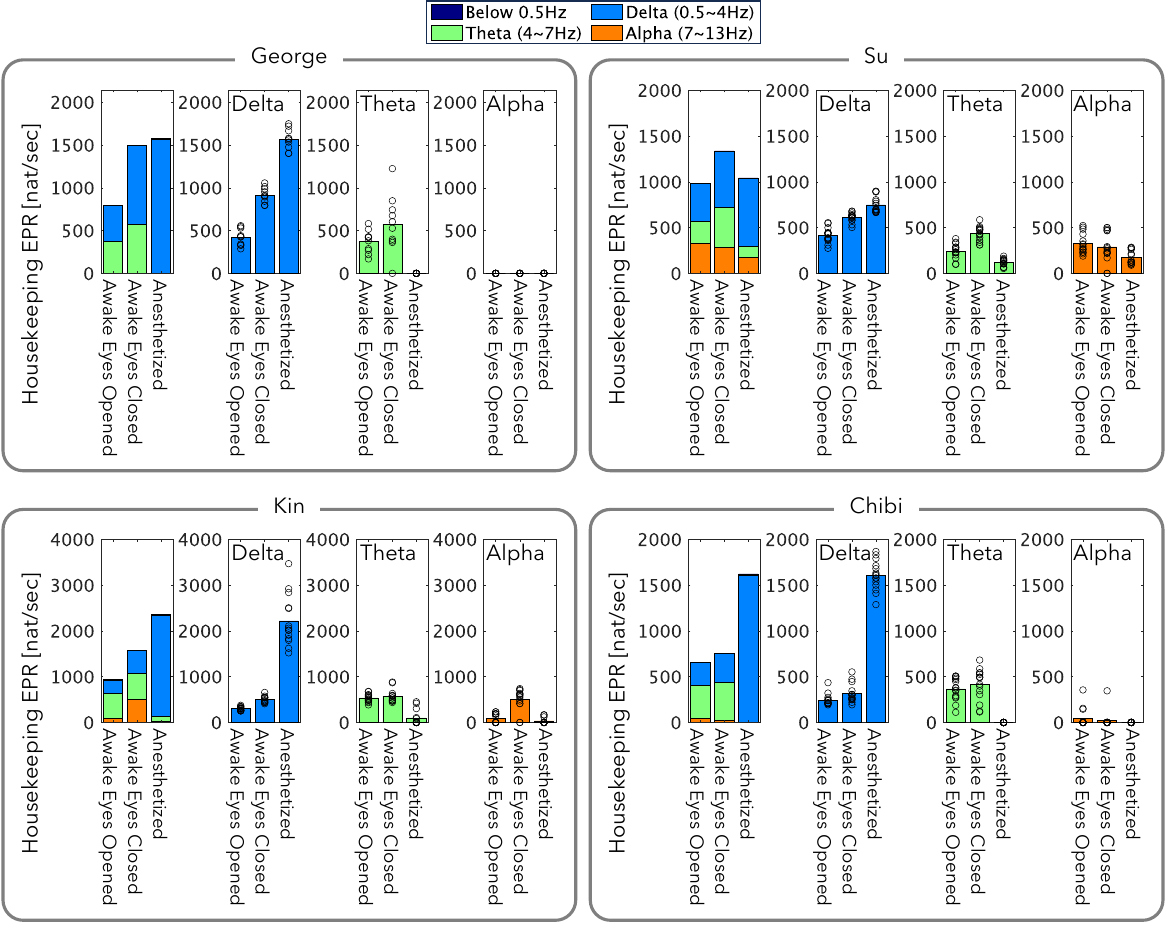}
\caption{
The temporal stability of the contributions of each frequency band to the housekeeping entropy production rate (EPR) was robust across all individual monkeys (George, Su, Kin and Chibi).  
The contribution to the housekeeping entropy production rate from the delta band $\sigma_t^{\mathrm{hk, Delta}}$, theta band $\sigma_t^{\mathrm{hk, Theta}}$ and alpha band $\sigma_t^{\mathrm{hk, Alpha}}$ [Eqs. (\ref{eq:def_sigma_delta}), (\ref{eq:def_sigma_theta}), and (\ref{eq:def_sigma_alpha})] are shown. 
The displayed values are the averages over time windows. The contributions to the housekeeping entropy production rate from the delta band $\sigma_t^{\mathrm{hk, Delta}}$, theta band $\sigma_t^{\mathrm{hk, Theta}}$, and alpha band $\sigma_t^{\mathrm{hk, Alpha}}$ are respectively shown. The bar plots represent the averaged values over time windows. Each circle represents the value of each time window. The plots for Kin show the same figure as Fig. \ref{fig:time_variatiom}b-e. 
}
\label{fig:time_variation_all-monkey}
\end{figure*}

\bibliography{apssamp}

\end{document}